\documentclass[graybox]{svmult}

\usepackage{type1cm}        
\usepackage{makeidx}         
\usepackage{graphicx}        
\usepackage{multicol}        
\usepackage[bottom]{footmisc}

\usepackage{algorithm}
\usepackage{algpseudocode}
\usepackage{comment}

\usepackage{newtxtext}       %
\usepackage[varvw]{newtxmath}       

\usepackage{tikz}
\usepackage{pgfplots}
\usetikzlibrary{intersections, pgfplots.fillbetween}
\usepackage{subcaption}
\usepackage{hyperref}
\hypersetup{
    colorlinks=true,        
    linkcolor=blue,         
    citecolor=blue,          
    urlcolor=black, 
}

\makeindex             

\newcommand{\ba}{\mathbf{a}}
\newcommand{\bb}{\mathbf{b}}
\newcommand{\bA}{\mathbf{A}}
\newcommand{\bd}{\mathbf{d}}
\newcommand{\e}{\mathsf{e}}
\newcommand{\jj}{\mathsf{j}}

\newcommand{\bH}{\mathbf{H}}
\newcommand{\bR}{\mathbf{R}}
\newcommand{\bx}{\mathbf{x}}
\newcommand{\bn}{\mathbf{n}}
\newcommand{\bff}{\mathbf{f}}
\newcommand{\bw}{\mathbf{w}}
\newcommand{\bW}{\mathbf{W}}

\newcommand{\by}{\mathbf{y}}
\newcommand{\bI}{\mathbf{I}}
\newcommand{\bu}{\mathbf{u}}

\newcommand{\bp}{\mathbf{p}}

\newcommand{\bPsi}{\boldsymbol{\Psi}}
\newcommand{\bmu}{\boldsymbol{\mu}}
\newcommand{\bSigma}{\boldsymbol{\Sigma}}
\newcommand{\bOmega}{\boldsymbol{\Omega}}
\newcommand{\blambda}{\boldsymbol{\lambda}}

\def\bzero{\boldsymbol{0}}
\def\bone{\boldsymbol{1}}

\newcommand{\Herm}{\mathsf{H}}
\newcommand{\Trans}{\mathsf{T}}
\newcommand{\x}{\mathsf{x}}
\newcommand{\y}{\mathsf{y}}
\newcommand{\z}{\mathsf{z}}
\newcommand{\bigo}{\mathcal{O}}

\def\LOS{\mathrm{LOS}}
\def\nLOS{\mathrm{nLOS}}
\def\tx{\mathrm{tx}}
\def\rx{\mathrm{rx}}

\def\sinc{\mathrm{sinc}}
\def\iid{\mathrm{iid}}
\def\PL{\mathrm{PL}}
\def\ris{\mathrm{ris}}
\def\bs{\mathrm{bs}}
\def\ilm{\mathrm{ilm}}
\def\uc{\mathrm{uc}}
\def\FF{\mathrm{FF}}
\def\NF{\mathrm{NF}}
\def\qNF{\mathrm{qNF}}
\def\Diag{\mathrm{Diag}}
\def\Rank{\mathrm{Rank}}
\newcommand{\Tr}{\mathrm{Tr}}

\def\Cset{\mathbb{C}}
\def\Rset{\mathbb{R}}
\def\Betaset{\hat{\boldsymbol{\beta}}}
\newcommand{\Ex}{\mathbb{E}}
\newcommand{\kk}{\kappa}

\def\sCN{\mathcal{CN}}
\def\Mset{\mathcal{M}}
\def\Aset{\mathcal{A}}
\def\Pset{\mathcal{P}}
\def\Uset{\mathcal{U}}
\def\Qset{\mathcal{Q}}

\newcommand{\vahid}[1]{{\color{DarkGreen}VJ: #1}}
\newcommand{\mohamadreza}[1]{{\color{blue}MD: #1}}
\definecolor{DarkGreen}{RGB}{0,150,0}
\definecolor{olivegreen}{HTML}{029A46}

\usepackage{tikz}
\usepackage{pgfplots}
\usetikzlibrary{plotmarks}
\usetikzlibrary{shapes,decorations.markings,fit,calc,patterns}
\usetikzlibrary{spy}
\usepgfplotslibrary{external}
\usetikzlibrary{external}
\newtheorem{lem}{Lemma}
\newtheorem{remk}{Remark}

\usepackage{IEEEtrantools} 
\usepackage[noadjust]{cite} 

\begin{document}

\title*{Far- versus Near-Field RIS Modeling and Beam~Design}
\author{Mohamadreza Delbari\orcidID{0000-0002-4768-5874} \\
George C. Alexandropoulos\orcidID{0000-0002-6587-1371} \\ Robert Schober\orcidID{0000-0002-6420-4884}  \\
Vahid Jamali\orcidID{0000-0003-3920-7415}}
\authorrunning{M. Delbari, G. C. Alexandropoulos, R. Schober, and V. Jamali}
\institute{Mohamadreza Delbari \at Technical University of Darmstadt, Darmstadt, Germany \email{mohamadreza.delbari@tu-darmstadt.de}
\and George C. Alexandropoulos \at National and Kapodistrian University of Athens, Athens, Greece \email{alexandg@di.uoa.gr}
\and Robert Schober \at Friedrich-Alexander University Erlangen-Nürnberg (FAU), Erlangen, Germany \email{robert.schober@fau.de}
\and Vahid Jamali \at Technical University of Darmstadt, Darmstadt, Germany \email{vahid.jamali@tu-darmstadt.de}}

%
%
\maketitle


\abstract{In this chapter, we investigate the mathematical foundation of the modeling and design of reconfigurable intelligent surfaces (RIS) in both the far- and near-field regimes. More specifically, we first present RIS-assisted wireless channel models for the far- and near-field regimes, discussing relevant phenomena, such as line-of-sight (LOS) and non-LOS links, rich and poor scattering, channel correlation, and array manifold. Subsequently, we introduce two general approaches for the RIS reflective beam design, namely optimization-based and analytical, which offer different degrees of design flexibility and computational complexity. Furthermore, we provide a comprehensive set of simulation results for the performance evaluation of the studied RIS beam designs and the investigation of the impact of the system parameters.}

\section{Introduction}
\label{sec: Introduction}

Reconfigurable intelligent surfaces (RISs) have been extensively investigated in the recent literature as a promising technology for introducing reconfigurability into the wireless channel, promoting the concept of  smart radio environments \cite{renzo2019smart,wu2019towards,yu2021smart,EURASIP_RIS_all}. RISs consist of passive programmable sub-wavelength elements,
so-called unit-cells or meta-atoms, that can change the properties of an impinging electromagnetic (EM) wave while reflecting it. By the joint optimization of the RIS unit-cells, advanced EM manipulation functionalities, such as anomalous reflection, beam focusing, and beam splitting can be achieved. These features have been exploited in the context of wireless communications for, e.g., realizing virtual line-of-sight (LOS) connections and improving the link budget, increasing the rank of the wireless channel, implementing over-the-air modulation, and enabling secure communications \cite{pan2022overview,Space_shift_keying_RIS,yu2020robust}. 

Both theoretical studies \cite{najafi2020physics,bjornson2020power} and recent experimental proof-of-concept implementations of RIS-assisted wireless communications \cite{tang2020wireless,dai2020reconfigurable} have suggested that for passive RISs to have a significant impact on the wireless channel, their electrical size must be extremely large, in fact, much larger than the size of typical active transmit/receive arrays. While the extremely large electrical dimension of RISs enables the realization of exciting features, such as precise three-dimensional (3D) beam focusing, it also introduces new challenges for their efficient modeling and optimization.  In particular, as the RIS dimension increases, the far-field distance increases, which implies that near-field models become essential in characterizing RIS-assisted wireless channels. 
Moreover, for extremely large purely reflective RISs, the transmitter (Tx)-RIS and RIS-receiver (Rx) channel matrices contain a huge number of channel coefficients, which makes the conventional approach of optimizing the RIS based on directly estimating the channel matrices infeasible due to the entailed large estimation overhead~\cite{Tsinghua_RIS_Tutorial}. A viable approach here is to devise near-field RIS reflection beams with a tuneable beamwidth that can adapt themselves to the accuracy of the estimated channel parameters (e.g., the locations of the Tx, Rx, and scattering objects) and to the affordable reconfiguration overhead \cite{jamali2022lowtozero,Alexandropoulos2022Near}. 

\textbf{Chapter contributions:} The focus of this book chapter is on the modeling and beam design of RIS-assisted wireless systems in both the near- and far-field regimes. To this end, first, the background on the mathematical characterization of near- and far-field regimes is concisely presented in Section~\ref{sec: far-near-field}. Therein, a sub-region is introduced within the near-field regime, called the quadratic near-field, where the wavefront phase-change across the observation plane can be approximated by a quadratic function (unlike the linear phase-change in the far field and the general non-linear phase variation due to the spherical propagation in the near-field). Subsequently, in Section~\ref{sec: System model}, the channel models for the far- and near-field regimes are introduced, discussing relevant phenomena such as LOS and non-LOS links, rich and poor scattering, channel correlation, and array manifold. In Section~\ref{sec: Approaches for beam design}, the design of RIS reflective beams, i.e., the RIS phase-shift configurations, is investigated. To this end, first, some key considerations for the channel state information (CSI) and performance metric used for the beam design are presented. Then, two general approaches for RIS beam design, namely optimization-based and analytical, are introduced for both the far-field and near-field regimes. In general, the optimization-based RIS designs yield a high-quality beam design at the expense of high computational complexity, whereas the analytical solutions to the RIS beam design problem offer less flexibility, but are more insightful and are easily scalable to extremely large RISs. In Section~\ref{sec: Simulation result}, we provide a comprehensive set of simulation results for the performance evaluation of the studied designs and the investigation of the impact of the various system parameters. Finally,  Section~\ref{sec:conclusions} summarizes the content of this chapter and draws some useful conclusions.   

\textbf{Related literature:}
Several works have studied RIS-assisted wireless systems in the near-field regime \cite{bjornson2021primer,haghshenas2023parametric,ramezani2022near,cui2021near,dardari2021nlos,abu2021near,liu2023near}. For instance, the modeling of near-field channels was the focus in \cite{bjornson2021primer} and channel estimation in the near-field regime was investigated in \cite{haghshenas2023parametric}. The authors in \cite{ramezani2022near} considered the problem of near-field beamforming and showed that data multiplexing is possible even in LOS channels. Near-field wideband beamforming for extremely large antenna arrays has been studied in \cite{cui2021near,Xu_DMA_2022}. Furthermore, the authors in \cite{dardari2021nlos,abu2021near} derived algorithms for localization for large RISs under the near-field assumption. Finally, the recent papers \cite{HMIMO_survey,liu2023near} provide comprehensive tutorial reviews on near-field communications. 

\textit{Notation:} Bold capital and small letters are used to denote matrices and vectors, respectively.  $(\cdot)^\Trans$, $(\cdot)^\Herm$, $\Rank(\cdot)$, and $\Tr(\cdot)$ denote the transpose, Hermitian, rank, and trace of a matrix, respectively. Moreover, $\Diag(\bA)$ is a vector that contains the main diagonal entries of matrix $\bA$, and $\ba\cdot\bb$ denotes the inner product of two vectors. $|a|$ denotes the absolute value of complex number $a$, $\|\ba\|$ represents the Euclidean norm of vector $\ba$, whereas $\|\bA\|_*=\sum_i \sigma_i$ and $\|\bA\|_2=\max_i \sigma_i$ denote the respectively nuclear and spectral norms of a Hermitian matrix $\bA$, where $\sigma_i,\,\,\forall i$, are the singular values of $\bA$. Furthermore, $\bzero_n$ and $\bone_n$ denote column vectors of size $n$ whose elements are all zeros and all ones, respectively, and $\mathbf{I}_n$ is the $n\times n$ identity matrix. $[\bA]_{m,n}$ and $[\ba]_{n}$ denote the element in the $m$th row and $n$th column of matrix $\bA$ and the $n$th entry of vector $\ba$, respectively. $\Rset$ and $\Cset$ represent the sets of real and complex numbers, respectively, $\jj$ is the imaginary unit, and $\Ex\{\cdot\}$  represents expectation. $\mathcal{CN}(\bmu,\bSigma)$ denotes a complex Gaussian random vector with mean vector $\bmu$ and covariance matrix $\bSigma$. Finally, $\bigo(\cdot)$ and $o(\cdot)$ represent the big-O and little-o notations, respectively.



\section{Background on Far- versus Near-field Regimes}
\label{sec: far-near-field}


The electrical field at a distance $z$ from a point source can be modelled as \cite{dardari2020,bjornson2021primer}:
\begin{equation}
\label{eq: electrical field}
    E = \frac{\jj\eta\e^{-\jj\frac{2\pi}{\lambda}z}}{2\lambda z}\Big(1+\frac{\jj}{2\pi z/\lambda}-\frac{1}{(2\pi z/\lambda)^2}\Big),
\end{equation}
where $\lambda$ is the wavelength and $\eta$ denotes the impedance of free space. Hence, its absolute value can be obtained as follows:
\begin{equation}
    |E| = \frac{\eta}{2\lambda z}\Big(1-\frac{1}{(2\pi z/\lambda)^2}+\frac{1}{(2\pi z/\lambda)^4}\Big).
\end{equation}
If the distance is smaller than the wavelength (i.e., $z<\lambda$), the last two terms in parentheses cannot be neglected. In other words, when the electrical field is measured very close to the source, the amplitude variation is significant. This region is referred to as the \textit{reactive near-field}. However, if the distance between the observation point and the source is larger than the wavelength (i.e., $z>\lambda$), the amplitude variations can be neglected, but the phase variations (including the wavefront curvature) are not necessarily negligible. In the remainder of this chapter, we focus on the distance beyond the reactive near-field, called the \textit{radiative near-field}. Moreover, for brevity, in the following, we refer to the radiative near-field simply as near-field. 

\begin{figure}[t]
	\centering
        \includegraphics[width=0.6\textwidth]{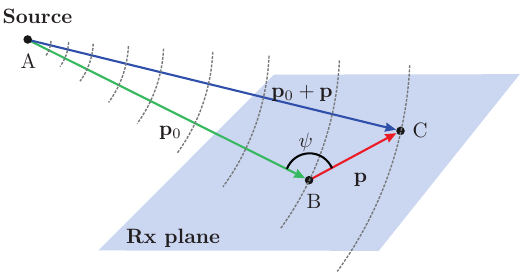}
    \caption{The change of wavefront across the Rx plane. A point source is located at A, the center of the Rx plane is located at B, and C is an arbitrary point on the Rx plane. Function $\Delta\phi(\bp_0,\bp)$, defined in \eqref{eq:phase_diff}, quantifies the change in the phase of the electric field at point C with respect to the phase at the Rx center B.}
    \label{fig:phase}
\end{figure}

We are interested in studying how the phase of the electric field changes across the Rx plane, see Fig.~\ref{fig:phase}. Let the source be at the center of the coordinate system and $\bp_0$ denotes the center of the Rx plane. The difference of the phase of the electric field  at an arbitrary point on the Rx surface, characterized by vector $\bp$ from the center of the Rx, with respect to (w.r.t.) the phase at the Rx center is given by
\begin{align}\label{eq:phase_diff}
    \Delta\phi(\bp_0,\bp) = \kappa \left(\|\bp_0+\bp\| - \|\bp_0\|\right),
\end{align}
where $\kappa=\frac{2\pi}{\lambda}$ denotes the wave number. The following lemma provides a useful result for distinguishing between the near- and far-field regimes.

\begin{lem}\label{lemma:phase_expand}
Assuming $\|\bp_0\|\gg \|\bp\|$, the change in the phase of the electric field across the Rx plane w.r.t. to the Rx center can be expanded as
\begin{align} \label{eq:phase_expand}
	\Delta\phi(\bp_0,\bp) = \kappa
		p_0\Big[ \cos(\psi)\tilde{p}+\frac{\sin^2(\psi)}{2}\tilde{p}^2 - \frac{\cos(\psi)\sin^2(\psi)}{2} \tilde{p}^3+o(\tilde{p}^4) \Big],
\end{align}
where $\tilde{p}=\frac{p}{p_0}$, $p_0=\|\bp_0\|$, $p=\|\bp\|$, and $\psi$ is the angle between vectors $\bp_0$ and $\bp$.
\end{lem}
\begin{IEEEproof}
    We start by rewriting $\|\bp_0+\bp\|^2$ as follows:
    \begin{align} \label{eq:innerproduct}
			\|\bp_0+\bp\|^2 & = (\bp_0+\bp)\cdot (\bp_0+\bp) \nonumber\\
   & = \|\bp_0\|^2+2\bp_0\cdot\bp+\|\bp\|^2 \nonumber\\
   & = p_0^2+2\cos(\psi)p_0p+p^2.
\end{align}
Therefore, we have
\begin{align}
    \|\bp_0+\bp\| = p_0 \sqrt{1+2\cos(\psi)\tilde{p}+\tilde{p}^2}.
    \label{eq:sumsofp}
\end{align}  
Note that, for most practical scenarios, $p_0\gg p$ holds since the RIS-transceiver distance is larger than the dimension of RIS. Therefore, we expand $\|\bp_0+\bp\|$ around $\tilde{p}=\frac{p}{p_0}=0$. To this end, we use the following Taylor series expansion \cite{wolfram2023}:
\begin{align}\label{eq:Taylor}
    \sqrt{1+x} = 1+\frac{x}{2}-\frac{x^2}{8}+\frac{x^3}{16}+o(x^4).
\end{align}  
Substituting $x=2\cos(\psi)\tilde{p}+\tilde{p}^2$ into \eqref{eq:Taylor}, we obtain 
\begin{align}
    &\sqrt{1+2\cos(\psi)\tilde{p}+\tilde{p}^2} \nonumber \\
    &= 1+\cos(\psi)\tilde{p}+\frac{\sin^2(\psi)}{2}\tilde{p}^2 - \frac{\cos(\psi)\sin^2(\psi)}{2} \tilde{p}^3+o(\tilde{p}^4).
\end{align} 
Substituting the above result into \eqref{eq:sumsofp} results in \eqref{eq:phase_diff} leading to \eqref{eq:phase_expand} which concludes the proof.
\end{IEEEproof}

Lemma~\ref{lemma:phase_expand} reveals how the phase of the electric field varies across the Rx surface in terms of the dominant linear, quadratic, and cubic terms of $\tilde{p}$. We use these terms to mathematically define the different regions.

\begin{trailer}{Far- vs. Near-Field Regions}
Far-field refers to the regime where the curvature of the wavefront is negligible, i.e., only the linear term in \eqref{eq:phase_expand} is non-negligible. A pragmatic condition for defining where the far field begins is to assume that the phase error caused by neglecting the quadratic term does not exceed $\frac{\pi}{8}$. This leads to  
\begin{align}\label{eq:FF_derivation}
    \kappa p_0 \frac{\sin^2(\psi)}{2}\tilde{p}^2 
    \overset{(a)}{\leq} \frac{\pi p^2}{\lambda p_0} 
    \overset{(b)}{\leq} \frac{\pi D^2}{4\lambda p_0}
    \leq \frac{\pi}{8},
\end{align} 
where $D$ is the largest dimension of the Rx, inequality $(a)$ follows from $|\sin(\psi)|\leq 1$, and inequality $(b)$ follows from $p\leq D/2$. The distance $p_0$ that meets the last inequality in \eqref{eq:FF_derivation} is called the far-field distance and is denoted by $d_\FF$:  
\begin{align}\label{eq:FF_distance}
    d_\FF = \frac{2D^2}{\lambda}.
\end{align}
The far-field distance is also known as the Rayleigh distance \cite{liu2023near}. For distances  $p_0<d_\FF$, the curvature of the wavefront cannot be neglected along the Rx surface, which we refer to as near field in this chapter. 
\end{trailer}

\begin{figure}[t]
	\centering
        \includegraphics[width=0.7\textwidth]{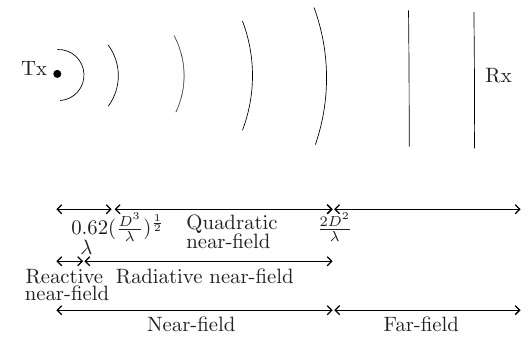}
    \caption{Illustration of different propagation regimes as a function of the distance to a point source.}
    \label{fig:near-far-field}
\end{figure}

\newpage
\begin{trailer}{Quadratic Near-Field Sub-Region}
Within the near field, we can distinguish a sub-region where the phase variation across the Rx plane can be modeled by a quadratic function of $p$. To obtain the condition for this region, we assume that the cubic term in \eqref{eq:phase_expand} is negligible (i.e., less than $\pi/8$). This leads to 
\begin{align}\label{eq:qNF_derivation}
    \kappa p_0 \frac{\cos(\psi)\sin^2(\psi)}{2} \tilde{p}^3
    \overset{(a)}{\leq}  \frac{\pi D^3}{12\sqrt{3}\lambda p_0^2}
    \leq \frac{\pi}{8},
\end{align} 
where inequality $(a)$ follows from $\cos(\psi)\sin^2(\psi)\leq \frac{2}{3\sqrt{3}}$ \cite{wolfram2023} and $p\leq D/2$. We refer to the distance $p_0$ that meets the last inequality in \eqref{eq:qNF_derivation} as the quadratic near-field distance, which is denoted by $d_\qNF$:  
\begin{align}\label{eq:qNF_distance}
    d_\qNF = \sqrt{\frac{2D^3}{3\sqrt{3}\lambda}}\approx 0.62 \sqrt{\frac{D^3}{\lambda}}.
\end{align}
The quadratic near-field distance is related to the Fresnel distance \cite{liu2023near}. In Section~\ref{sec:analytic_solutions}, we will exploit the quadratic phase change in this region to develop analytical solutions for the RIS phase configuration.  For distances  $p_0<d_\qNF$, the higher order terms or the full spherical wave propagation need to be considered.
\end{trailer}

\begin{figure}[t]

\begin{minipage}{1\textwidth}
	\centering
    \includegraphics[width=0.8\textwidth]{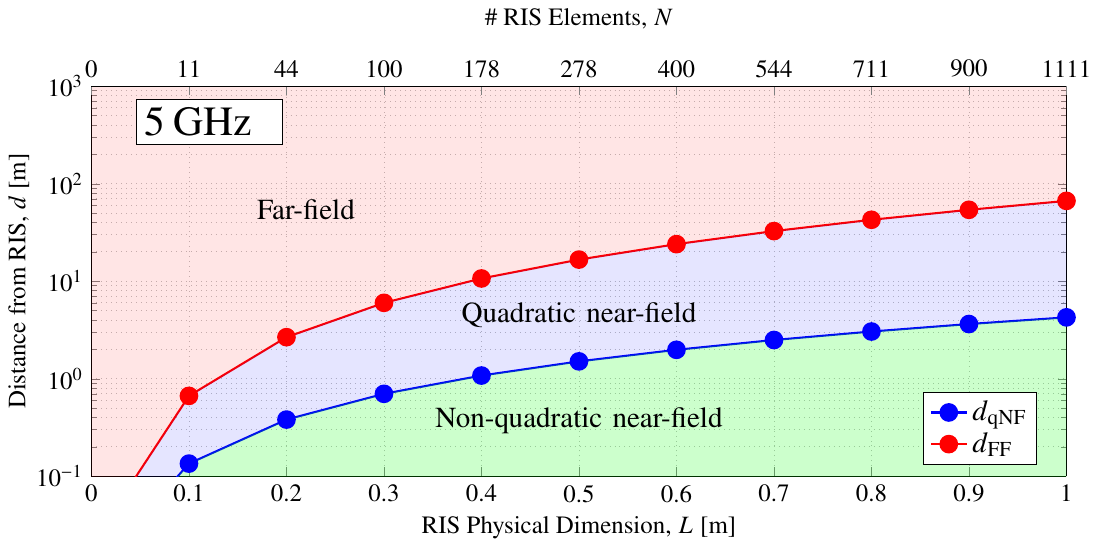}
 \end{minipage}
 
 \vspace{0.3cm}
 
\begin{minipage}{1\textwidth}
	\centering
        \includegraphics[width=0.8\textwidth]{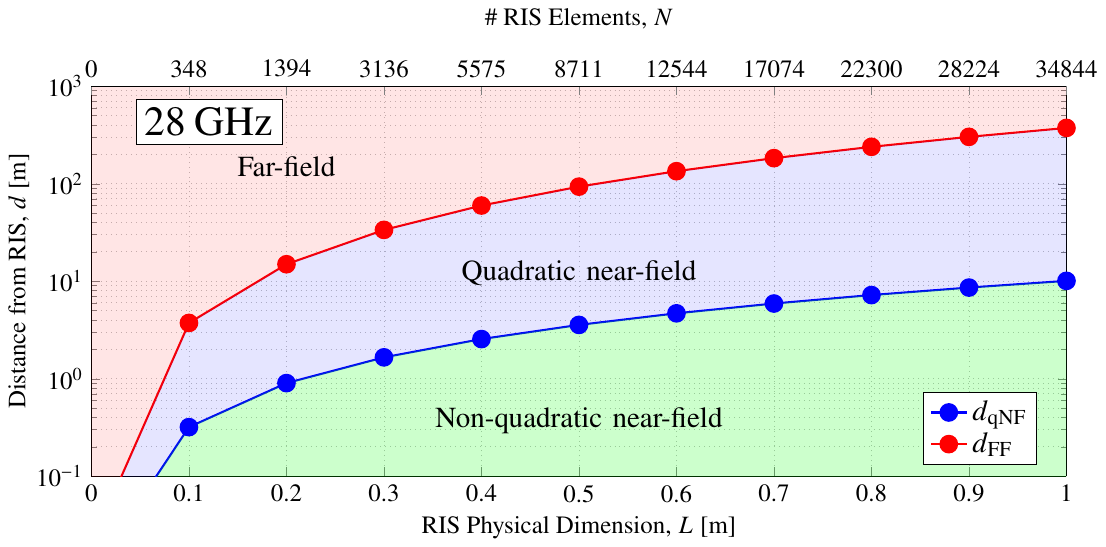}
\end{minipage}

     \caption{Far-field and quadratic near-field distances of a square RIS with horizontal and vertical lengths $L$ (i.e., $D = \sqrt{2}L$) for carrier frequencies of $5$ GHz (upper figure) and $28$ GHz (lower figure). The required number $N$ of unit cells is also shown for half-wavelength element spacing.}
	\label{fig: near-quad-field}
\end{figure}

The far- and near-field regimes, as well as the quadratic near-field sub-region, are illustrated in Fig.~\ref{fig:near-far-field}. Moreover, Fig.~\ref{fig: near-quad-field} quantitatively shows the far-field distance $d_\FF$ and the quadratic near-field distance $d_\qNF$ for a square RIS with horizontal and vertical length $L_\y = L_\z \triangleq L$ (i.e., $D = \sqrt{L_\y^2+L_\z^2}=\sqrt{2}L$) for carrier frequencies of $5$~GHz (upper figure) and $28$~GHz (lower figure). As an example, assuming $L=0.5$~m, we obtain $d_\FF=16$~m and $d_\qNF=1.5$~m for $5$~GHz carrier frequency and $d_\FF=93$~m and $d_\qNF=3.5$~m for $28$~GHz carrier frequency. As can be seen from Fig.~\ref{fig: near-quad-field}, the quadratic near-field sub-region covers most of the near-field region that is of practical interest.



\section{System and Channel Models}
\label{sec: System model}

\begin{figure}[t]
\begin{minipage}{0.5\columnwidth}
	\centering 
        \includegraphics[width=1\textwidth]{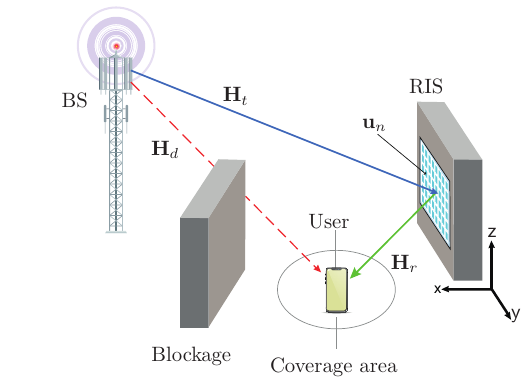}
\end{minipage}
\begin{minipage}{0.5\columnwidth}
	\centering 
        \includegraphics[width=1\textwidth]{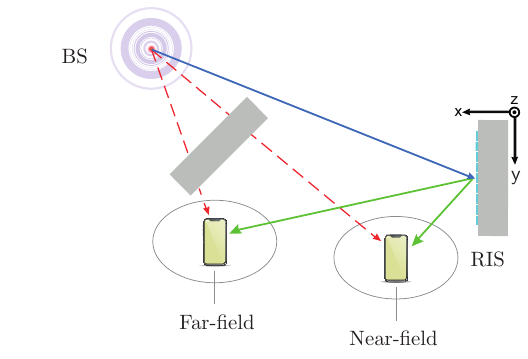}
\end{minipage}
    \caption{Schematic illustration of an RIS-assisted downlink wireless communication system. The left-hand side figure shows a 3D model whereas the right-hand side figure presents a 2D model of the same scenario. We adopt both 2D and 3D models for our simulation results in Section~\ref{sec: Simulation result}, whereby a detailed discussion on the definitions and values of the system parameters are provided.}
    \label{fig:system model}
\end{figure}

We consider a narrowband downlink communication system comprising a base station (BS) with $N_t$ antennas, an RIS with $N$ unit cells, and a mobile user (MU) with $N_r$ antennas, as depicted in Fig.~\ref{fig:system model}. The received signal can be modeled as
\begin{equation}\label{Eq:IRSbasic}
	\by = \big(\bH_d+\bH_r \bOmega \bH_t \big)\bx +\bn,
\end{equation}
where $\bx\in\Cset^{N_t}$ is the transmit signal vector satisfying $\Ex\{\bx^\Herm\bx\}\leq P_t$ with $P_t$ being the maximum transmit power, $\by\in\Cset^{N_r}$ is the received signal vector, and  $\bn\in\Cset^{N_r}$ represents the additive white Gaussian noise (AWGN), i.e., $\bn\sim\sCN(\bzero_{N_r},\sigma_n^2\bI_{N_r})$, where $\sigma_n^2$ is the noise power. Moreover, $\bH_{d}\in\Cset^{N_r\times N_t}, \bH_t\in\Cset^{N\times N_t}$, and $\bH_{r}\in\Cset^{N_r\times N}$ denote the BS-MU, BS-RIS, and RIS-MU channel matrices, respectively. Furthermore, $\bOmega\in\Cset^{N\times N}$ is a diagonal matrix with main diagonal entries $\Omega_n\e^{\jj\omega_n}$, where  $\omega_n$ (with $n=1,2,\ldots,N$) is the phase shift applied by the $n$th RIS unit-cell element and $\Omega_n$ is the unit-cell factor~\cite{huang2019reconfigurable}. The value of $\Omega_n$ in principle depends on the wavelength $\lambda$, unit-cell area $A_\uc$, and the incident and reflection angles. For simplicity, we assume a constant unit-cell factor $\Omega_n=\Omega\triangleq4\pi A_\uc\lambda^{-2}$  \cite{najafi2020physics} (see \cite{Wu2021} for a discussion on the dependency of the amplitude of the reflection coefficient on the desired phase-shift value).

In the following subsections, we define the channel models for the far- and near-field regimes, respectively. For notational simplicity, we drop subscripts $d$, $t$, and $r$, and explain the channel models for a general matrix $\bH\in\Cset^{N_\rx\times N_\tx}$ corresponding to $N_\tx$ transmit antenna elements and $N_\rx$ receive antenna elements, and wherever necessary, we explicitly refer to the  BS-UE, BS-RIS, and RIS-UE channels.

\subsection{Far-Field Channel Model}
\label{far-field}
It is usually desirable to deploy the RIS such that there exist LOS links in the both BS-RIS and RIS-UE channels. In this case, the BS-RIS and RIS-UE channels can be modeled by Rician fading  \cite{wu2019intelligent,Jamali2023impact}, i.e.,
\begin{IEEEeqnarray}{ll}\label{Eq:Rician}
	\bH = \sqrt{\frac{K}{1+K}}\bH^\LOS + \sqrt{\frac{1}{1+K}}\bH^{\nLOS},
\end{IEEEeqnarray}
where $\bH^\LOS$ and $\bH^\nLOS$ are the LOS and non-LOS components of $\bH$, respectively, and $K$ denotes the Ricean factor that determines the relative power of the LOS component compared to the non-LOS components of the channel. If the direct BS-MU channel is blocked, the LOS component does not exist for the BS-MU link leading to $K=0$ for this channel. Similarly, if the RIS is deployed in a rich scattering environment, the dominant LOS component can be modeled by assuming $K\to\infty$ for the BS-RIS and RIS-MU links~\cite{huang2019reconfigurable}.  

\textbf{LOS Component:} As discussed in Section~\ref{sec: far-near-field}, in the far-field regime, the variation of the phase of the electric field across the Rx array is a linear function of the distance with the Tx. Therefore, the LOS channel matrix $\bH^\LOS$ depends only on the deployment of the antenna arrays and the relative angles of the transmit and receive arrays. This leads~to
\begin{IEEEeqnarray}{ll}\label{Eq:LOSff}
	\bH^\LOS = c\, \ba_{\rx}(\bPsi_\rx)\ba_{\tx}^\Herm(\bPsi_\tx),
\end{IEEEeqnarray}
where $c$ is the channel attenuation factor of the LOS link, and $\ba_{\tx}(\cdot)\in\Cset^{N_\tx}$ and $\ba_{\rx}(\cdot)\in\Cset^{N_\rx}$ denote the transmit and receive array steering vectors, respectively. Moreover, $\bPsi_\tx=(\theta_\tx,\phi_\tx)$ is the Tx angle-of-departure (AoD), i.e., the direction of the Rx defined in the Tx array coordinate system, where $\theta_\tx$ and $\phi_\tx$ denotes the elevation and azimuth angles, respectively. Similarly, $\bPsi_\rx=(\theta_\rx,\phi_\rx)$ denotes the Rx angle-of-arrival (AoA), i.e., the direction of the Tx defined in the Rx array coordinate system. The array steering vector depends on the array manifold, i.e., the positions of the antenna elements, denoted by $\bu_n\in\Rset^3$, with $n=1,\dots,N$ and $N$ being the total number of array elements, and is given~by \cite{Jamali2023impact}:
\begin{IEEEeqnarray}{ll}\label{Eq:steering}
	\ba(\bPsi) = \left[\e^{\jj\kappa\bd^\Trans\!(\bPsi)\bu_1},\dots,\e^{\jj\kappa\bd^\Trans\!(\bPsi)\bu_N}\right]^\Trans,
\end{IEEEeqnarray}
where $\bd(\bPsi)\in\Rset^3$ is a unit vector pointing in the direction of $\bPsi$. Depending on the specific adopted array manifold and the choice of coordinates system, \eqref{Eq:steering} can be often further simplified. For instance, assuming a uniform planar array (UPA) located in the $\y-\z$ plane and consisting of $N_\y$ and $N_\z$ antenna elements (i.e., $N_\y N_\z=N$) spaced by $d_\y$ and $d_\z$ along the $\y$- and $\z$-axes, indexed by $n_\y=0,\dots,N_\y-1$ and $n_\z=0,\dots,N_\z-1$, respectively, we obtain $\bu_{(n_\y,n_z)}=[0,n_\y d_\y, n_\z d_\z]^\Trans$  and $\bd(\bPsi)=[\cos(\theta) \cos(\phi), \cos(\theta) \sin(\phi), \sin(\theta)]^\Trans$, where the first antenna element $(n_\y,n_\z)=(0,0)$ is assumed to be located at the origin. Therefore, $[\ba(\bPsi)]_n$ corresponding to antenna element $n$ (parameterized by $(n_\y,n_z)$ in the UPA) is given by
\begin{IEEEeqnarray}{ll}\label{Eq:UPA}
	[\ba(\bPsi)]_n = \e^{\jj\kappa \left[d_\y \cos(\theta) \sin(\phi) n_\y + d_\z \sin(\theta) n_\z \right]}.
\end{IEEEeqnarray}

\textbf{Non-LOS Component:} The non-LOS channel originates from the scattering objects in the environment and its structure significantly depends on whether the scattering is rich (as is often the case in sub-6 GHz communication systems) or poor (as is the case in millimeter wave (mmWave) and THz communication systems). We will investigate both cases in the following.

\textit{Rich Scattering:} In general, in rich scattering environments with ideally an infinite number of channel paths, the non-LOS channel matrix can be modeled using transmit and receive correlation matrices denoted by $\bR_\tx\in\Cset^{N_\tx\times N_\tx}$  and $\bR_\rx\in\Cset^{N_\rx \times N_\rx}$, respectively  \cite{bjornson2020rayleigh,Jamali2023impact}. More specifically, for UPAs, it has been shown in \cite{Jamali2023impact} that under two assumptions, namely A1 (a normalization for channel power) and A2 (uniform AoAs and AoDs), the non-LOS channel matrix can be rewritten as
	\begin{IEEEeqnarray}{ll}\label{Eq:matGaussianGenerate}
		\bH^\nLOS = \bar{\bR}_\rx\bH^\iid \bar{\bR}_\tx,
	\end{IEEEeqnarray}
	where $\bar{\bR}_s\in\Cset^{N_s\times N_s}$, $s\in\{\tx,\rx\}$, is obtained from the decomposition $\bR_s = \bar{\bR}_s\bar{\bR}_s^\Trans$ and the entries of $\bH^\iid\in\Cset^{N_\rx\times N_\tx}$ are independent and identically distributed (i.i.d.) complex Gaussian random variables distributed as $\sCN(0,\sigma_c^2)$, where $\sigma_c^2$ determines the power of the non-LOS component of the channel. The  transmit and receive correlation matrices are given by \cite[Lemma~1]{Jamali2023impact}
	\begin{IEEEeqnarray}{ll}\label{Eq:CorrelationMatrix}
	[\bR_s]_{m,n} = \sinc\big(\kappa\|\bu_{s,m}-\bu_{s,n}\|\big),
	\quad s\in\{\tx,\rx\},
\end{IEEEeqnarray}
where $\sinc(x)=\frac{\sin(x)}{x}$ is the sinc function and $\bu_{s,n}$ denotes the position of each $n$th antenna at node $s\in\{\tx,\rx\}$. The special case of i.i.d. Rayleigh fading of non-LOS channel is obtained by setting $\bR_s=\bI_{N_s}$, $s\in\{\tx,\rx\}$.

\textit{Poor Scattering:} In poor scattering environments, e.g., for mmWave and THz communications, the non-LOS channel is characterized by a limited number of channel paths, where in the far-field regime, each path is characterized by its AoA and AoD, and is modeled similar to \eqref{Eq:LOSff}. Depending on the roughness of the reflecting object, it may generate not only a single reflection path, but a collection of scattered paths that are spatially confined. To model this behavior, the channel is characterized by a collection of channel scattering objects each generating a number of channel paths that are closely spaced in the angular domain. This leads to the following model:\cite{jaeckel2017quadriga}
\begin{IEEEeqnarray}{ll}\label{Eq:nLOSlowrank}
	\bH^\nLOS \!=\! \frac{1}{\sqrt{VR}} \sum_{v=1}^V \sum_{r=1}^R g^{(v)}\e^{\jj\psi^{(v,r)}}\!\! \ba_{\rx}(\bPsi_\rx^{(v,r)})\ba_{\tx}^\Herm(\bPsi_\tx^{(v,r)}), \quad\,\,
\end{IEEEeqnarray}
where $\bPsi_\tx^{(v,r)}$ ($\bPsi_\rx^{(v,r)}$) is the AoD from the Tx (AoD on the Rx) for the $r$th sub-path in the $v$th cluster, $V$ and $R$ are the numbers of clusters and sub-paths in each cluster, respectively, $g^{(v)}$ is the path attenuation coefficient for the $v$th scatter cluster, and $\psi^{(v,r)}$ is the respective phase of the $r$th sub-path in cluster $v$.

\subsection{Near-Field Channel Model}
\label{near-field}

Similar to the far-field, the channel in the near-field may have both LOS and non-LOS components, and hence, can similarly be modeled by Rician fading as in \eqref{Eq:Rician}, although, in the near-field, it is more likely that the LOS component is strongly dominant for the BS-RIS and RIS-MU links. In the following, we discuss the near-field channel model for both the LOS and non-LOS components.

\textbf{LOS Component:} As discussed in Section~\ref{sec: far-near-field}, in the near field, the phase change across the Rx array depends on the distance between the Tx and Rx antennas. This leads to the following LOS near-field channel matrix:
\begin{IEEEeqnarray}{ll}\label{Eq:LOSnf}
	\big[\bH^\LOS\big]_{m,n} = c\, \e^{\jj\kk\|\bu_{\rx,m}-\bu_{\tx,n}\|},
\end{IEEEeqnarray}
where $c$ is the channel attenuation of the LOS path (similar to \eqref{Eq:LOSff} for the far field).

\textbf{Non-LOS Component:} The structure of the non-LOS channel matrices depends on whether or not the scattering environment is rich. For a rich scattering environment, if the scattered paths arrive at the Rx from all directions, a similar channel model to that in \eqref{Eq:matGaussianGenerate} can be adopted. However, in the near field, some reflection/scattering objects that are close to the Tx/Rx or have a large extent (e.g., walls or floor) may have a significant contribution to the channel matrix, and hence, be dominant. In the following, we introduce near-field channel models for two types of reflection/scattering objects. 

\begin{figure}[t]
\begin{minipage}{0.5\columnwidth}
	\centering 
	\centering
        \includegraphics[width=1\textwidth]{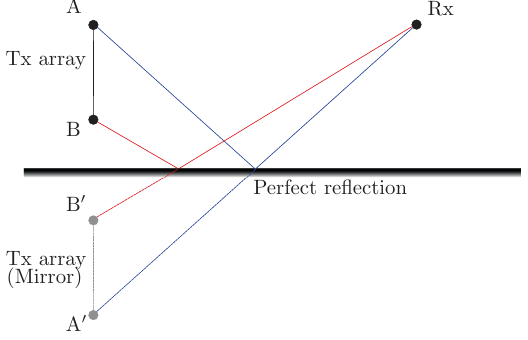}
\end{minipage}
\begin{minipage}{0.5\columnwidth}
	\centering 
	\centering
        \includegraphics[width=1\textwidth]{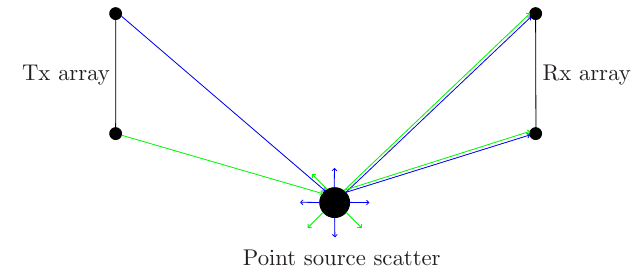}
\end{minipage}
	
    \caption{Illustration of perfect reflection (the left-hand side figure) and point-source scattering (the right-hand side figure) in the near-field region.}
    \label{fig:point-source-scatter}
\end{figure}

\textit{Perfect Reflection Model:} An important contribution to the non-LOS channel is the reflection from walls and floors, which typically have very large extents compared to the Tx/Rx arrays. Using image theory and geometric optics \cite[Ch. 4]{a2005antenna}, the end-to-end channel matrix resulting from perfect reflection by a surface with an infinite extent is given by
\begin{IEEEeqnarray}{ll}\label{Eq:reflector_nLOSscatter}
	\big[\bH_s^\nLOS\big]_{m,n} = c_r\, \e^{\jj\kk\|\bu^{\rm vrt}_{\rx,m}-\bu_{\tx,n}\|} = c_r\, \e^{\jj\kk\|\bu_{\rx,m}-\bu^{\rm vrt}_{\tx,n}\|},
\end{IEEEeqnarray}
where $\bu^{\rm vrt}_{\tx,n}$ ($\bu^{\rm vrt}_{\rx,n}$) is the virtual mirror image of the Tx (Rx) at the reflecting surface (see Fig.~\ref{fig:point-source-scatter} where A$^\prime$ and B$^\prime$ are the virtual mirror image points A and B, respectively) and $c_r$ is the end-to-end channel attenuation. Note that, based on the perfect reflection model,  $c_r$ is proportional to  $\frac{1}{\|\bu^{\rm vrt}_{\rx}-\bu_{\tx}\|}=\frac{1}{\|\bu_{\rx}-\bu^{\rm vrt}_{\tx}\|}$, where $\bu^{\rm vrt}_{\tx}$ and $\bu^{\rm vrt}_{\rx}$ are the centers of the mirror images of the Tx and Rx arrays, respectively.

\textit{Point Scattering Model:} In this model, it is assumed that the scattering object is small and scatters the wave in all directions, see Fig.~\ref{fig:point-source-scatter}. The end-to-end channel matrix resulting from such a point-source scatterer is given by
\begin{IEEEeqnarray}{ll}\label{Eq:point_nLOSscatter}
	\big[\bH_s^\nLOS\big]_{m,n} = c_s\, \e^{\jj\kk(\|\bu_{\rx,m}-\bu_{s}\|+\|\bu_s-\bu_{\tx,n}\|)},
\end{IEEEeqnarray}
where $\bu_{s}$ is the position of the $s$th point scatter and $c_s$ is the end-to-end path loss. Due to the double path-loss, $c_s$ is proportional to $\frac{1}{\|\bu_{\rx}-\bu_{s}\|\times\|\bu_s-\bu_{\tx}\|}$, where $\bu_{\tx}$ and $\bu_{\rx}$ are the centers of the Tx and Rx arrays, respectively. This result implies that point-source scattering has a negligible contribution unless the distance of the scatterer to either the Tx or the Rx is small.

While practical scattering/reflecting objects are neither point sources nor perfect reflectors, these idealistic models provide insights into how the near-field channel can be characterized under extreme scenarios, and hence, can be instrumental for the RIS beam design.



\section{RIS Reflective Beam Design}
\label{sec: Approaches for beam design}
In this section, we focus on RIS beam design, i.e., the design of RIS phase configuration $\bw \triangleq [\e^{\jj \omega_1},\dots,\e^{\jj \omega_N}]^\Trans$. To this end, we first discuss the CSI, design goal, and performance metric used for the beam design. Subsequently, we introduce two design approaches, one based on an optimization formulation and another analytical, for both the far- and near-field regimes. 


\subsection{Design Considerations}\label{Sec:design}

\subsubsection{CSI Requirement for RIS Beam Design}
While for performance analysis it is generally desirable to employ a channel model that is accurate and accounts for all relevant phenomena in the wireless channel (e.g., LOS, non-LOS, far- vs. near-field effects, channel correlation, poor vs. rich scattering, etc.; see Section~\eqref{sec: System model}), it is crucial to base the system design on a type of CSI whose acquisition is practically viable in real-time. The type of required CSI is particularly important for configuring almost passive RISs, which compared to typical arrays deployed at Txs and Rxs, are expected to be extremely large\footnote{In fact, for RISs to be able to realize a sufficient link budget, their size \textit{must} be extremely large to compensate for the inherent double path loss of RIS-generated virtual channels \cite{najafi2020physics}.}. Therefore, while using \textit{full CSI}, e.g., $\bH_t$ and $\bH_r$, for the of almost passive RISs leads to maximum performance, the estimation of these large matrices may not be feasible in practice~\cite{Tsinghua_RIS_Tutorial}. In contrast, one may base the RIS design on the estimation of a few of the \textit{channel parameters} that play a dominant role in shaping $\bH_t$ and $\bH_r$. For instance, for LOS channels, the estimates of the AoAs/AoDs to/from the Rx/Tx (in the case of the far field) or the locations of the Rx/Tx (in the case of the nearfield) may be exploited for the RIS phase configuration. Similarly, for the non-LOS components, the estimates of a few dominant paths can be exploited to further improve the RIS beam design, i.e., the AoAs/AoDs for the far field or the scattering locations for the near field.

In addition, for RIS-assisted wireless systems, one can distinguish between the CSI employed for RIS reconfiguration and the CSI used for optimizing the rest of the communication system, e.g., the precoder at the BS \cite{jamali2022lowtozero}. The CSI for RIS reconfiguration can be acquired much less frequently, since the channel parameters, such as the location of the MU, generally vary much more slowly than the coherence time of the channels $\bH_t$ and $\bH_r$. However, once the RIS is configured, i.e., $\bw$ is fixed, its impact on the channel is transparent to the BS and MUs, and the end-to-end channel $\bH_{e2e}=\bH_d+\bH_r^\Herm \bOmega \bH_t$ can be estimated more frequently based on the channel coherence time, as is done in conventional communication systems without RISs. From this perspective, the RIS does not adapt itself to the small-scale fading, but more generally, generates an end-to-end channel $\bH_{e2e}$ with favorable statistical features (e.g., sufficient link budget, large rank, etc.).

\subsection{Beam Design Objective}

Motivated by the above discussion, in the remainder of this section, we focus on RIS beam designs that require knowledge of only the dominant AoAs/AoDs or the locations of the BS, MU, or dominant scatters. In particular, we develop beam designs for \textit{anomalous reflection/focusing} with \textit{tunable beamwidths}. The beamwidth tunability is beneficial for the following reasons:
\begin{itemize}
    \item \textbf{Parameter estimation error:} The estimation of the AoAs/AoDs or MU locations is prone to error. A RIS beam with tuneable beamwidth can adjust itself to the quality of parameter estimation, which leads to a more robust design. 
    \item \textbf{Overhead of RIS reconfiguration:} The larger the beamwidth is, the less frequently the RIS has to be reconfigured\footnote{The reduced overhead comes at the expense of a reduced received power, which suggests a fundamental tradeoff between the performance and overhead of RIS phase reconfiguration, see \cite{Jamali2021quadratic} for a detailed discussion.}. Therefore, a tunable beamwidth implies a tunable overhead for RIS phase reconfiguration.
\end{itemize}

We note that the proposed RIS beam design can be implemented either \textit{online} or \textit{offline}. In the former case, the channel parameters are estimated and fed to the beam synthesis algorithm for real-time beam design. In the latter case, an over-complete codebook of RIS phase-shift configurations is designed offline for a discretized set of channel parameters, where the best phase-shift configuration is selected online based on the estimated channel parameters. Note that these two approaches require different implementations for the RIS control channel~\cite{RIS_control}.
\\

As a performance metric, we consider the generalized radar cross section (GRCS) $g_{\rm ris}$ defined in \cite{najafi2020physics}, which is used to determine the free-space end-to-end path loss of the RIS-enabled wireless link, denoted by $\PL_{\rm ris}$, according to \cite[Lemma~1]{najafi2020physics}\footnote{In \eqref{eq:pathloss}, we normalized \cite{najafi2020physics}'s GRCS definition by $\frac{\lambda}{\sqrt{4\pi}}$ in order to get a unitless quantity.}:
\begin{IEEEeqnarray}{ll}\label{eq:pathloss}
\PL_\ris \triangleq \PL_{t} \PL_{r} |g_\ris|^2,
\end{IEEEeqnarray}
where $\PL_t$ and $\PL_r$ are the free-space path losses of the BS-RIS and RIS-MU  links, respectively. In the far-field regime, $g_\ris(\bPsi_t,\bPsi_r|\bw)$ determines the power of the reflected wave along any AoD $\bPsi_r$ for an incident wave coming from any given AoA $\bPsi_t$, assuming that the RIS is configured according to phase configuration $\bw$. In the near-field regime, $g_\ris(\bu_t,\bu_r|\bw)$ specifies the power of the reflected wave at any location $\bu_r$ for an incident wave originating from any given location $\bu_t$, assuming that the RIS is configured according to phase configuration $\bw$.

In the following, we introduce two approaches, namely an optimization-based and an analytical, for the RIS reflective beam design.

\subsection{Optimization-based RIS Beam Design}

We first formulate optimization problems for the design of the RIS phase-shift configuration for the far- and near-field regions, respectively. Subsequently, we present a solution for these problems and analyze its computational complexity.

\subsubsection{Problem Formulation for the Far-Field Regime}
We develop an RIS phase configuration design that provides coverage for all reflection angles $\bPsi_r\in\Aset_r$ for any (unknown) incident angle $\bPsi_t\in\Aset_t$. The sizes of sets $\Aset_t$ and $\Aset_r$ depend on the accuracy of the estimated AoAs and AoDs as well as the affordable overhead; see Section~\ref{Sec:design}. In particular, we consider the following optimization problem:
\begin{IEEEeqnarray}{ll}
    \text{P1:}\quad&~\underset{\bw,\gamma}{\max} \,\,\,\, \gamma \nonumber\\
    \text {s.t.} &~\text{C1:}  \quad \big| g_\ris\big(\bPsi_t,\bPsi_r|\bw\big) \big|^2 \geq\gamma,\quad\forall (\bPsi_t,\bPsi_r)\in\Aset_t\times\Aset_r\nonumber\\
    &~\text{C2:} \quad \big|[\bw]_{n} \big|=1, \quad\forall n=1, \dots, N,
\end{IEEEeqnarray}
where $\gamma$ is an auxiliary optimization variable. Based on the discussion in Section~\ref{far-field}, the RIS GRCS can be obtained in the far-field regime as follows: 
\begin{IEEEeqnarray}{ll}\label{eq:GRCSfar}
   g_\ris\big(\bPsi_t,\bPsi_r|\bw\big) & = \ba_{\ris}^\Herm(\bPsi_r) \bOmega \ba_{\ris}(\bPsi_t)  \nonumber\\
   & = \Omega \sum_{n=1}^N \e^{\jj\kk(\bd(\bPsi_t)-\bd(\bPsi_r))^\Trans\bu_n + \jj \omega_n} = \bff_\FF^\Herm(\bPsi_t,\bPsi_r) \bw,
\end{IEEEeqnarray}
where $\ba_{\ris}(\cdot)\in\Cset^{N}$ is the RIS array steering vector, $\bu_n$ is the location of the $n$th element of the RIS, and $\bff_\FF(\bPsi_t,\bPsi_r) \in\Cset^N$ is given by
\begin{IEEEeqnarray}{ll}
   [\bff_\FF(\bPsi_t,\bPsi_r)]_n = \Omega \e^{-\jj\kk(\bd(\bPsi_t)-\bd(\bPsi_r))^\Trans\bu_n}
   \overset{(a)}{=} \Omega \e^{-\jj\kk(\beta_\y n_\y + \beta_\z n_\z)}.
\end{IEEEeqnarray}
In this expression, equality $(a)$ holds for an UPA with the steering vector in \eqref{Eq:UPA}, where 
\begin{IEEEeqnarray}{ll}\label{Eq:ByBz}
   \beta_\y=d_\y (\cos(\theta_t) \sin(\phi_t)-\cos(\theta_r) \sin(\phi_r)), \IEEEyesnumber\IEEEyessubnumber\\
   \beta_\z=d_\z (\sin(\theta_t)-\sin(\theta_r)). \IEEEyessubnumber
\end{IEEEeqnarray}

\subsubsection{Problem Formulation for the Near-Field Regime}

Analogous to the beam design in the far-field region, for the near-field, our objective is to develop an RIS phase configuration that provides coverage for all observation locations $\bu_r\in\Uset_r$ for an incident wave originating from a source at any (unknown) location $\bu_t\in\Uset_t$. Again, the sizes of sets $\Uset_t$ and $\Uset_r$ depend on the accuracy of the estimated source and observation locations as well as the affordable overhead; see Section~\ref{Sec:design}. Specifically, we formulate the following optimization problem:
\begin{IEEEeqnarray}{ll}
    \text{P2:}\quad&~\underset{\bw,\gamma}{\max}\,\,\,\, \gamma \nonumber\\
    \text {s.t.} &~\text{C1:}  \quad \big|g_\ris\big(\bu_t,\bu_r|\bw\big)\big|^2 \geq\gamma,\quad\forall (\bu_t,\bu_r)\in\Uset_t\times\Uset_r\nonumber\\
    &~\text{C2:} \quad \big|[\bw]_{n} \big|=1, \quad\forall n=1, \dots, N.
\end{IEEEeqnarray}
Using the results presented 
in Section~\ref{far-field}, the RIS GRCS in the near-field regime can be obtained as follows:
\begin{IEEEeqnarray}{ll}\label{eq:GRCSnear}
   g_\ris\big(\bu_t,\bu_r|\bw\big) = \Omega \sum_{n=1}^N \e^{\jj\kk(\|\bu_t-\bu_n\| + \|\bu_r-\bu_n\|) + \jj \omega_n} = \bff_\NF^\Herm(\bu_t,\bu_r) \bw,
\end{IEEEeqnarray}
where $\bff_\NF(\bu_t,\bu_r) \in\Cset^N$ is given by
\begin{IEEEeqnarray}{ll}
   [\bff_\NF(\bu_t,\bu_r)]_n = \Omega \e^{-\jj\kk(\|\bu_t-\bu_n\| + \|\bu_r-\bu_n\|)}.
\end{IEEEeqnarray}

\subsubsection{Beam Design Solution}

As can be seen from P1 and P2, both problems have the same functional form in the optimization variables $\bw$ and $\gamma$, namely:
\begin{IEEEeqnarray}{ll}
    \text{P3:}\quad&~\underset{\bw,\gamma}{\max} \,\,\,\, \gamma \nonumber\\
    \text {s.t.} &~\text{C1:}  \quad \big|\bff_q^\Herm \bw\big|^2 \geq\gamma,\quad\forall q\in\Qset\nonumber\\
    &~\text{C2:} \quad \big|[\bw]_{n} \big|=1, \quad\forall n=1, \dots, N,
\end{IEEEeqnarray}
where for the far field, $\bff_q=\bff_\FF(\bPsi_t,\bPsi_r)$ and set $\Qset$ contains all desired AoAs and AoDs $(\bPsi_t,\bPsi_r)\in\Aset_t\times\Aset_r$, whereas for the near field, $\bff_q=\bff_\NF(\bu_t,\bu_r)$ and set $\Qset$ contains the location of all the desired source and observation points $(\bu_t,\bu_r)\in\Uset_t\times\Uset_r$.


The main difficulty in solving P3 is the non-convex unit-modulus constraint C2, which is inherent to the RIS passiveness assumption. There are various approaches to cope with C2, including semidefinite relaxation (SDR) with Gaussian randomization \cite{wu2019intelligent}, alternating optimization (AO) among the phase shift of individual unit-cells \cite{wu2019beamforming}, penalty-based method \cite{wu2020joint}, successive convex approximation (SCA) \cite{yang2020intelligent}, manifold optimization \cite{yu2020optimal,RIsopt2023},
deep learning-based method \cite{feng2020deep}, and alternating direction method of multipliers (ADMM) and majorization-minimization (MM) methods \cite{liu2022joint}. In the following, we use a method from  \cite{Ghanem2023Optimization,yu2020robust} that is based on the reformulation and relaxation of the unit-modulus constraint. In particular, by defining a new optimization variable $\bW=\bw\bw^\Herm$, C1 and C2 can be equivalently rewritten as:
\begin{IEEEeqnarray}{ll}
    \IEEEyesnumber\IEEEyessubnumber*
    &~\widebar{\text{C1}}:\quad\,\,\, \bff_q^\Herm\bW\bff_q\geq \gamma, \quad \forall q\in\Qset \\*
    &~\widebar{\text{C2a}}:\quad \Diag(\bW)=\textbf{1}_N \IEEEyessubnumber\\
    &~\widebar{\text{C2b}}:\quad \bW\succeq 0 \IEEEyessubnumber\\
    &~\widebar{\text{C2c}}:\quad \Rank(\bW)=1. \IEEEyessubnumber
\end{IEEEeqnarray}

The equivalent constraints are still non-convex due to the rank constraint in $\widebar{\text{C2c}}$. To cope with this issue, we rewrite $\widebar{\text{C2c}}$ equivalently as follows \cite{Yu2020power}:
\begin{IEEEeqnarray}{ll}
    &~\widebar{\text{C2c}}:\quad \|\bW\|_*-\|\bW\|_2 \leq 0.
\end{IEEEeqnarray}
Although $\widebar{\text{C2c}}$ is still non-convex, it is the difference of two convex functions. This feature has been used in \cite{Ghanem2023Optimization,yu2020robust} to obtain a sub-optimal solution based on the penalty method and SCA. First, using the penalty method, we rewrite P3 as follows:
\begin{IEEEeqnarray}{ll}\label{Eq:P3bar}
    \widebar{\text{P3}}:\quad&~\underset{\bW,\gamma}{\max}~\gamma -\eta  (\|\bW\|_*-\|\bW\|_2) \nonumber\\
    \text {s.t.} &~\widebar{\text{C1}},\,\, \widebar{\text{C2a}},\,\, \widebar{\text{C2b}},
\end{IEEEeqnarray}
where $\eta\geq 0$ is a factor for penalizing the violation of $\widebar{\text{C2c}}$. It has been shown in \cite[Proposition 2]{yu2020robust} that, for sufficiently large values of $\eta$, constraint $\widebar{\text{C2c}}$ is enforced, and hence, $\widebar{\text{P3}}$ becomes equivalent to P3. The last step is applying SCA to $\|\bW\|_2$ using the following Taylor series approximation bound evaluated at initial point $\|\bW^{(i)}\|_2$:
\begin{IEEEeqnarray}{ll}\label{Eq:Taylor}
    \|\bW\|_2\, &\geq \|\bW^{(i)}\|_2 
    + \Tr\Big(\nabla_{\bW}^\Herm (\|\bW\|_2)\big|_{\bW=\bW^{(i)}} \times (\bW-\bW^{(i)})\Big) \nonumber\\
    &=\|\bW^{(i)}\|_2 
    + \Tr\Big(\blambda_{\max}(\bW^{(i)})\blambda_{\max}^\Herm(\bW^{(i)})(\bW-\bW^{(i)})\Big),
\end{IEEEeqnarray}
where $\nabla_{\bW}(\cdot)$ denotes matrix differentiation w.r.t. $\bW$ and $\blambda_{\max}(\bA)$ is the eigenvector associated with the maximum eigenvalue of matrix $\bA$. Substituting the lower bound in \eqref{Eq:Taylor} into \eqref{Eq:P3bar}, we obtain the following approximation of P3: 
\begin{IEEEeqnarray}{ll}\label{Eq:P3tilde}
    \widetilde{\text{P3}}:\quad&~\underset{\bW,\gamma}{\max}~\gamma -\eta^{(i)}  \Big(\|\bW\|_*-\|\bW^{(i)}\|_2 
    - \Tr\Big(\blambda_{\max}(\bW^{(i)})\blambda_{\max}^\Herm(\bW^{(i)})(\bW-\bW^{(i)})\Big)\Big) \nonumber\\
    \text {s.t.} &~\widebar{\text{C1}},\,\, \widebar{\text{C2a}},\,\, \widebar{\text{C2b}}.
\end{IEEEeqnarray}
Optimization problem $\widetilde{\text{P3}}$ is convex since the objective
function is concave and the constraints span a convex set.
Hence, this problem can be efficiently solved by standard convex
optimization solvers, such as CVX \cite{cvx}. Algorithm~\ref{alg:SCA} summarizes
the main steps for solving P3 in an iterative manner,
where the solution of $\widetilde{\text{P3}}$ in iteration $i$ is used as the initial point for the next iteration $i + 1$ \cite{Ghanem2023Optimization}. The sequence of penalty factors $\eta^{(i)}$ is generated in a monotonically increasing manner to ensure constraint $\widebar{\text{C2c}}$ holds for large $i$. Thereby, Algorithm~\ref{alg:SCA} produces a sequence of improved feasible
points until convergence to a locally optimum point of P3. After convergence, the solution $\bW$ is rank-one, and hence, it can be obtained based on the eigenvalue decomposition of $\bW$, as shown in Algorithm~\ref{alg:SCA}. 

\begin{algorithm}[t]
\caption{Penalty- and SCA-based solution of P3 \cite{yu2020robust,Ghanem2023Optimization}}\label{alg:SCA}
\textbf{Initialization:}  Set initial matrix $\bW^{(0)}$, iteration index $i = 0$, maximum number of iterations $I_{\text{max}}$, $\alpha>1$, and penalty factors $\eta^{(0)} > 0$, and $\eta_{\text{max}}$.
\begin{algorithmic}
\Repeat
    \State Solve convex problem $\widetilde{\text{P3}}$ for given $\bW^{(i)}$ and store the intermediate solution $\bW$.
    \State Set $i = i + 1$ and update  $\bW^{(i)}=\bW$ as well as $\eta^{(i)} = \min(\alpha\eta^{(i-1)}, \eta_\text{max})$.
    \Until{Convergence or $i=I_{\text{max}}$}
    \State Set $\bw=\sqrt{\delta_{\max}}\blambda_{\max}(\bW)$ where $\delta_{\max}=\Tr(\bW)$.
\end{algorithmic}
\textbf{Output:} $\bw^{*}=\bw$.
\end{algorithm}

\subsubsection{Complexity Analysis}
Optimization problem $\widetilde{\text{P3}}$ is a semidefinite programming (SDP) problem. The computational complexity required per iteration
for solving an SDP with a numerical convex program solver
is given by $\bigo(mn^3 + m^2n^2 + m^3)$ \cite{bomze2010interior,najafi2019c}, where $m$ and $n$ denote the number of semidefinite cone constraints and the
dimension of the semidefinite cone, respectively. Thus, considering $m=|\Qset|$ and $n=N$ for problem $\widetilde{\text{P3}}$, the complexity order of Algorithm~\ref{alg:SCA} per iteration is given by
\begin{equation}
    \bigo\Big(|\Qset|N^3+|\Qset|^2N^2+|\Qset|^3\Big)\overset{(a)}{\approx}\bigo\Big(|\Qset|N^3\Big),
    \label{eq: complexity}
\end{equation}
where approximation $(a)$ assumes that $N\gg|\Qset|$. Therefore, the computational complexity of the proposed optimization-based method for RIS beam design grows cubically with the number of RIS unit-cells $N$, which may become prohibitive for extremely large RISs. Moreover, the efficiency of Algorithm~\ref{alg:SCA} relies on a good initial point, which, in principle, is difficult to obtain. In the following subsection, we present analytical solutions for the RIS beam design that are efficient to compute and can also serve as initial point for Algorithm~\ref{alg:SCA}.

\subsection{Analytical Beam Design}\label{sec:analytic_solutions}

In the following,  we introduce analytical solutions for the same objective as for P1 and P2 (namely anomalous reflection/focusing with tunable beamwidth) for the far- and near-field regimes, respectively.

\subsubsection{Far-Field Regime}

In \cite{Jamali2021quadratic}, an analytical solution was proposed for the RIS codebook design for arbitrary codebook sizes. To enable different codebook sizes, a so-called quadratic phase-shift profile was proposed, which we use in the following as the basis for RIS beam design with a tunable beamwidth covering $(\bPsi_t,\bPsi_r)\in\Aset_t\times\Aset_r$. To start, let us first assume $|\Aset_t|=|\Aset_r|=1$, which for the far-field regime, implies reflecting an incident plane wave from a given AoA $\bPsi_t$ to the desired AoD $\bPsi_r$, which leads to the following well-know beamforming solution:
\begin{IEEEeqnarray}{ll}\label{Eq:Linear}
   \omega_n = -\kk(\bd(\bPsi_t)-\bd(\bPsi_r))^\Trans\bu_n
   \overset{(a)}{=} -\kk(\beta_\y n_\y + \beta_\z n_\z),
\end{IEEEeqnarray}
where equality $(a)$ holds for an UPA, where $\beta_\y$ and $\beta_\z$ are given in \eqref{Eq:ByBz}, and each RIS unit cell index $n$ corresponds to a unique index pair $(n_\y,n_\z)$. In the following, we focus on the UPA, however, the proposed design is also applicable to other array manifolds. Note that the phase shift in \eqref{Eq:Linear} is a linear function in spatial variables $(n_\y,n_\z)$ and leads to the narrowest beam that the RIS can realize in the far field. Obviously, the design of a beam that covers $(\bPsi_t,\bPsi_r)\in\Aset_t\times\Aset_r$ demands a non-linear phase-shift profile, which in the simplest case, can be realized by a quadratic phase-shift solution~\cite{Jamali2021quadratic}, as follows:
\begin{IEEEeqnarray}{ll}\label{Eq:Quadratic}
   \omega_n = -\kk(\alpha_\y n_\y^2+\gamma_\y n_\y) -\kk(\alpha_\z n_\z^2+\gamma_\z n_\z),
\end{IEEEeqnarray}
where $\alpha_s$ and $\gamma_s$, $s\in\{\y,\z\}$ are given by
\begin{IEEEeqnarray}{ll}\label{Eq:AlphaGamma}
   \alpha_s = \frac{1}{2N_s}\left[\underset{(\bPsi_t,\bPsi_r)\in\Aset_t\times\Aset_r}{\max} \, \beta_s(\bPsi_t,\bPsi_r) - \gamma_s \right], \label{Eq:Alpha}\IEEEyesnumber\IEEEyessubnumber\\
   \gamma_s =  \underset{(\bPsi_t,\bPsi_r)\in\Aset_t\times\Aset_r}{\min} \, \beta_s(\bPsi_t,\bPsi_r), \label{Eq:Gamma} \IEEEyessubnumber
\end{IEEEeqnarray}
and $\beta_s(\bPsi_t,\bPsi_r)$ as a function of $\bPsi_t$ and $\bPsi_r$ is given in \eqref{Eq:ByBz}. In essence, the above design ensures that the gradient of the phase-shift profile (i.e., $-\kk\sum_{s\in\{\y,\z\}}(2\alpha_s n_s+\gamma_s)$) contains components for reflection of any $(\bPsi_t,\bPsi_r)\in\Aset_t\times\Aset_r$ across the RIS dimension, i.e., from $n_s=0$ to $n_s=N_s-1,\,\,s\in\{n_\y,n_\z\}$. Another useful interpretation of the quadratic phase-shift design in \eqref{Eq:Quadratic} is that $\gamma_s$ determines where the reflection begins, whereas $\alpha_s$ specifies the width of the reflected beam. Obviously, the linear phase-shift design in \eqref{Eq:Linear} can be obtained as a special case of the quadratic phase-shift design in \eqref{Eq:Quadratic} by setting $\alpha_s=0,\,\,s\in\{n_\y,n_\z\}$.

\subsubsection{Near-Field Regime}

The idea of the quadratic phase-shift design was extended in \cite{jamali2022lowtozero} to the design of a wide near-field illumination and in \cite{Alexandropoulos2022Near} to the design of a hierarchical near-field phase-shift codebook. Before applying the design in \cite{jamali2022lowtozero,Alexandropoulos2022Near} to the near-field illumination covering $(\bu_t,\bu_r)\in\Uset_t\times\Uset_r$, we start with the simple instructive case where  $|\Uset_t|=|\Uset_r|=1$, which implies focusing a spherical wave originated from a source located at $\bu_t$ to an observation point at $\bu_r$. This problem has the following well-known beam focusing solution: 
\begin{IEEEeqnarray}{ll}\label{Eq:Focusing}
   \omega_n \,&= -\kk \sum_{p\in\{t,r\}}\|\bu_p-\bu_n\| + \|\bu_p\| \nonumber\\
   &\overset{(a)}{=}-\kk \sum_{p\in\{t,r\}} \bar{\alpha}_p \|\bu_n\|^2+\bar{\gamma}_p \|\bu_n\| + \|\bu_p\|.
\end{IEEEeqnarray}
Equality $(a)$ in \eqref{Eq:Focusing} holds for the quadratic near-field sub-region (see Section~\ref{sec: far-near-field}), where parameters $\bar{\alpha}_p$ and $\bar{\gamma}_p,\,\,p\in\{t,r\}$, can be obtained from Lemma~\ref{lemma:phase_expand} as follows:
\begin{IEEEeqnarray}{ll} 
   \bar{\alpha}_p = \frac{\sin^2{\psi_{n,p}}}{2\|\bu_p\|}, \IEEEyesnumber\IEEEyessubnumber\\
   \bar{\gamma}_p = \cos(\psi_{n,p}), \IEEEyessubnumber
\end{IEEEeqnarray}
where $\psi_{n,p}$ is the angle between vectors $\bu_n$ and $\bu_p,\,\,p\in\{t,r\}$. As can be seen from \eqref{Eq:Focusing}, in the quadratic near-field sub-region, the phase shifts required for beam focusing can be approximated by a function that is quadratic in $\|\bu_n\|$, i.e., the distance between the center of the RIS to the $n$th RIS element.  Note that the term $-\kk\|\bu_p\|$ in \eqref{Eq:Focusing} is a constant phase shift for all unit cells, and hence does, not change the shape of the beam pattern; this implies that it can be dropped for beam focusing. Nonetheless, this term is required for the wide-beam near-field solution discussed in the following. 

We refer to the phase-shift in \eqref{Eq:Focusing} as $\omega_{\NF,n}(\bu_t,\bu_r)$ to explicitly show the dependency of this phase shift on $\bu_t$ and $\bu_r$. Let $\bu\in\Uset_{\ris}$ denote all the points on the RIS. To illuminate the entire desired area, one may partition the RIS $\Uset_{\ris}$ into sub-surfaces and split the targeted illumination area $\Uset_t\times\Uset_r$ into sub-regions, where each sub-surface on the RIS illuminates the centers of one sub-region. A general formulation of this problem is to develop a continuous mapping $\Mset:\Uset_{\ris}\rightarrow\Uset_t\times\Uset_r$ from the RIS surface space $\Uset_{\ris}$ to the space $\Uset_t\times\Uset_r$. Using this notation, we introduce the following phase-shift design:
\begin{IEEEeqnarray}{ll}\label{Eq:WideFocusing}
   \omega_n \,&= \omega_{\NF,n}(\Mset(\bu_n)).
\end{IEEEeqnarray}
In the following, we present an example of mapping $\Mset$. Let us assume the RIS is placed in the $\y-\z$ plane with its center located at $\bu_{\ris}$ and having size $L_\y\times L_\z$, the location of the source is known $|\Uset_t|=1$, and we are interested in illuminating a rectangular area in the $\x-\y$ plane with center $\bu_{c}$ and size $R_\x\times R_\y$. Then, assuming $\bu_n=\bu_{\ris}+[0,\y_n,\z_n]$ and $\bu_{c} = [\x_c,\y_c,\z_c]$, a possible mapping is \cite[Example~1]{jamali2022lowtozero}
\begin{IEEEeqnarray}{ll}\label{Eq:Mapping} 
   \Mset(\bu_n) = \bu_c + \left[\frac{R_\x}{L_\z}\z, \frac{R_\y}{L_\y}\y,0\right].
\end{IEEEeqnarray}
The idea behind the mapping suggested in \eqref{Eq:Mapping} is as follows. We assumed that the elements on the RIS along the $\y$-axis are responsible for covering the points along the $\y$-axis in the coverage area, whereas the elements on the RIS along the $\z$-axis are responsible for covering the points along the $\x$-axis in the coverage area. Therefore, for example, by continuously changing $\z$ on the RIS from $-\frac{L_\z}{2}$ to $\frac{L_\z}{2}$, all the points from $\x_c-\frac{R_\x}{2}$ to $\x_c+\frac{R_\x}{2}$ on the coverage area  are mapped. Similarly, by continuously varying $\y$ on the RIS from $-\frac{L_\y}{2}$ to $\frac{L_\y}{2}$, all the points from $\y_c-\frac{R_\y}{2}$ to $\y_c+\frac{R_\y}{2}$ in the coverage area are covered.

\section{Performance Evaluation}
\label{sec: Simulation result}

In this section, we first introduce the considered simulation setup. Subsequently, we present our simulation results and evaluate the performance of the proposed RIS phase configuration designs.

\subsection{Simulation Setup}
We adopt the system setup for coverage extension depicted in Fig.~\ref{fig:system model}. The center of the RIS array, the estimated center of the BS array, and the center of the illumination area are denoted by $\bu_{\bs}$, $\bu_\ris$, and $\bu_\ilm$, respectively, where by convention, we assume $\bu_{\ris}=[0,0,0]$ and define $d_t=\|\bu_{\bs}\|$ and $d_r=\|\bu_{\ilm}\|$. Moreover, we assume a multiple-antenna BS, a passive and purely reflective RIS with half-wavelength element spacing, and a single-antenna MU. For ease of presentation and to investigate specific features of the considered RIS beam design, we show results in 2D and 3D separately. Similarly, we provide results for free-space propagation (i.e., LOS) and multipath propagation environments (i.e., both LOS and non-LOS), respectively. The MU is randomly located within the targeted illumination area and the results provided are averaged over $100$ realizations of the MU location. Similarly, for the cases where the BS location is known up to a certain estimation error, the actual BS location is generated randomly within the error space around the estimated value. Furthermore, we use a carrier frequency of 28 GHz. The parameters used in Algorithm~\ref{alg:SCA} are as follows: $\eta^{(0)}=0.001$, $\alpha=5$, $\eta_{\max}=5000$, and $I_{\max}=10$. The remaining simulation parameters are provided in the description of each figure. 

The four algorithms presented in Section~\ref{Sec:design}, namely the optimization-based and analytical methods for the near- and far-field regimes are used for RIS beam design. To be able to compare different scenarios in a fair manner, we adopt the normalized GRCS $\bar{g}_\ris\triangleq|g_\ris|/g_{\max}$ as a performance metric with $g_\ris$ defined in \eqref{eq:pathloss} and the normalization factor $g_{\max} = \Omega N$ is the maximum attainable GRCS; see \eqref{eq:GRCSfar} and \eqref{eq:GRCSnear}. Since the exact locations of the BS and/or MU are unknown, we show the normalized GRCS for the worst case, i.e., we show
\begin{align}\label{eq:nGRCS}
\min_{q\in\Qset} \bar{g}^2_\ris(q) = \min_{q\in\Qset} \frac{|g_\ris(q)|^2}{g^2_{\max}}=\min_{q\in\Qset} \frac{|\bff_q^\Herm\bw|^2}{g^2_{\max}},
\end{align}
where $\Qset$ denotes the set of possible locations of the BS and MU. Moreover, we show simulation results in terms of signal-to-noise ratio (SNR) for a BS transmit power of $P_t=20$ dBm and a noise power of $\sigma_n^2=WN_0N_f$, where $N_0 = -174$ dBm/Hz is the noise power spectral density, $W = 20$ MHz is the signal bandwidth, and $N_f = 6$ dB is the Rx noise figure. Furthermore, we show results for benchmark schemes that employ full CSI, perfect beamforming, perfect focusing, random reflection, and specular reflection.

\begin{remk}
The MATLAB codes used to generate the simulation results in this section are publicly available online at \url{https://github.com/MohamadrezaDelbari/Far-versus-Near-Field-RIS-Modeling-and-Beam-Design}.
\end{remk}


\subsection{Simulation Results}

In the following, we first study the convergence behavior of Algorithm~\ref{alg:SCA} for the presented optimization-based RIS design. Subsequently, we investigate the resulting illumination beam pattern for both the far- and near-field regimes. For these results, we focus on a 2D setting and free-space propagation. Finally, we adopt a 3D setting with multipath propagation and evaluate the impact of non-LOS channel components on the performance of the proposed RIS reflective beam designs.

\begin{figure}[t]
     \begin{subfigure}{1\textwidth}
     \centering
         \includegraphics[width=\textwidth]{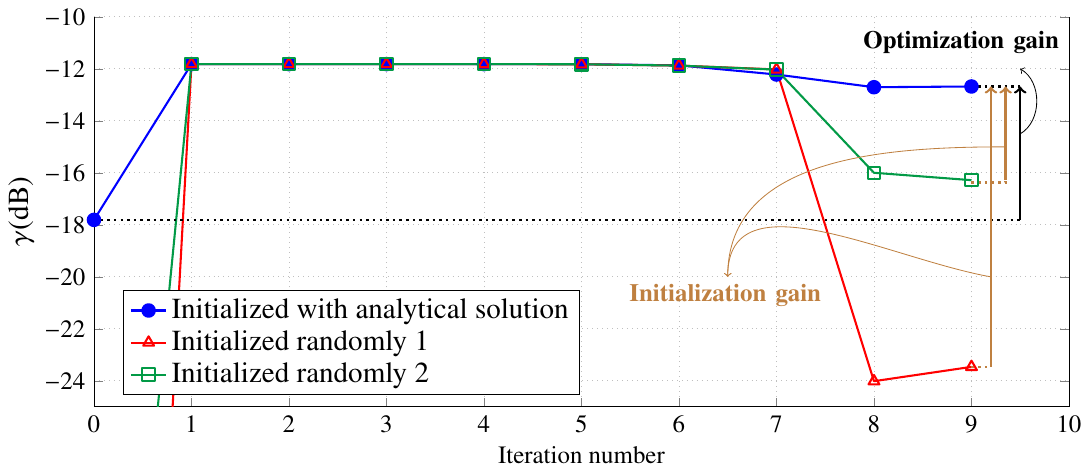}
         \caption{$\gamma=\min\limits_{q\in\Qset} \bar{g}^2_\ris(q)$ (dB) vs. the iteration number.}
        \label{fig: random_or_analytical_iteration_a}
   \end{subfigure}
        \begin{subfigure}{1\textwidth}
     \centering
         \includegraphics[width=\textwidth]{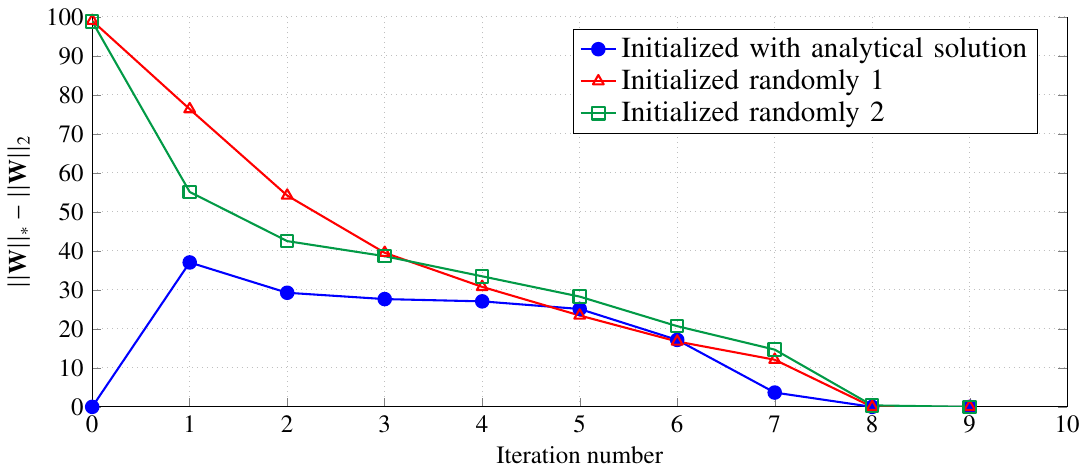}
         \caption{$\|\bW\|_*-\|\bW\|_2$ vs. the iteration number.}
        \label{fig: random_or_analytical_iteration_b}
   \end{subfigure}
   \caption{Convergence behavior of Algorithm~\ref{alg:SCA} for different initializations. Sub-plot (a) shows the normalized GRCS, whereas subplot (b) presents the rank constraint in $\overline{\text{C2c}}$. }
   \label{fig: random_or_analytical_iteration}  
\end{figure}


\textbf{Convergence Behavior:} In Fig.~\ref{fig: random_or_analytical_iteration}, we study the convergence behavior of  Algorithm~\ref{alg:SCA} for an example scenario in the far-field regime. The BS is located at $\bu_\bs=[30,80,0]$~m and the MU is randomly located within the area $\{(\x,\y,\z):-10~\text{m}\leq\x\leq 10~\text{m}, 60~\text{m}\leq\y\leq 80~\text{m}, \z=0\}$. Figure~\ref{fig: random_or_analytical_iteration_a} depicts the normalized GRCS\footnote{More specifically, when the rank constraint is relaxed, the normalized GRCS is obtained as $\min\limits_{q\in\Qset}\bff_q^\Herm\bW\bff_q / g^2_{\max}$, which becomes identical to \eqref{eq:nGRCS} after Algorithm~\ref{alg:SCA} converges to a feasible solution.} in \eqref{eq:nGRCS} as a function of the number of algorithmic iterations, whereas Fig.~\ref{fig: random_or_analytical_iteration_b} presents the rank constraint in $\widebar{\text{C2c}}$. Moreover, we show results for two initialization schemes, namely initialization with the analytical solution in \eqref{Eq:Quadratic} and random initializations. As can be seen from Fig.~\ref{fig: random_or_analytical_iteration_b}, the penalty method eventually enforces the rank constraint to hold regardless of the initialization strategy. For the case of initialization with the analytical solution, the rank constraint initially holds; however, it is relaxed by Algorithm~\ref{alg:SCA} to explore the space for a potentially better solution, which is indeed achieved leading to a $5$~dB performance gain for this example. As expected, when initialized randomly, the rank constraint is not met initially\footnote{Figure~\ref{fig: random_or_analytical_iteration_b} illustrates that the initial values of $\|\bW\|_*-\|\bW\|_2$ for both random initializations are approximately 100. This is because matrix $\bW$ contains $N\times N=10^4$ entries, and despite being randomly initialized, the randomness is averaged out leading to similar values for $\|\bW\|_*-\|\bW\|_2$.} and the value of the normalized GRCS is small as well. However, with sufficient iterations, Algorithm~\ref{alg:SCA} finds a feasible solution but the achievable normalized GRCS after convergence significantly depends on the initialization. Moreover, from Fig.~\ref{fig: random_or_analytical_iteration_a}, we observe that the value of the normalized GRCS is similar for the few first iterations regardless of the initialization, which implies that problem $\widebar{\text{P3}}$ with relaxed rank constraint (i.e., small penalty factor $\eta$) has multiple optimal points. Nonetheless, despite having similar normalized GRCS, the underlying solution $\bW$ is different (also apparent from Fig.~\ref{fig: random_or_analytical_iteration_b}), which leads to a considerably different normalized GRCS after convergence to a rank-one solution, i.e., $3$~dB and $10$~dB gains are achieved by initialization with the analytical solutions compared to the two cases of random initialization, respectively.  
Overall, these observations underline the importance of initialization and the usefulness of the proposed analytical solution for this purpose.


\begin{figure}
\centering
\begin{subfigure}{0.32\textwidth}
    \includegraphics[width=\textwidth,height=0.8\textwidth]{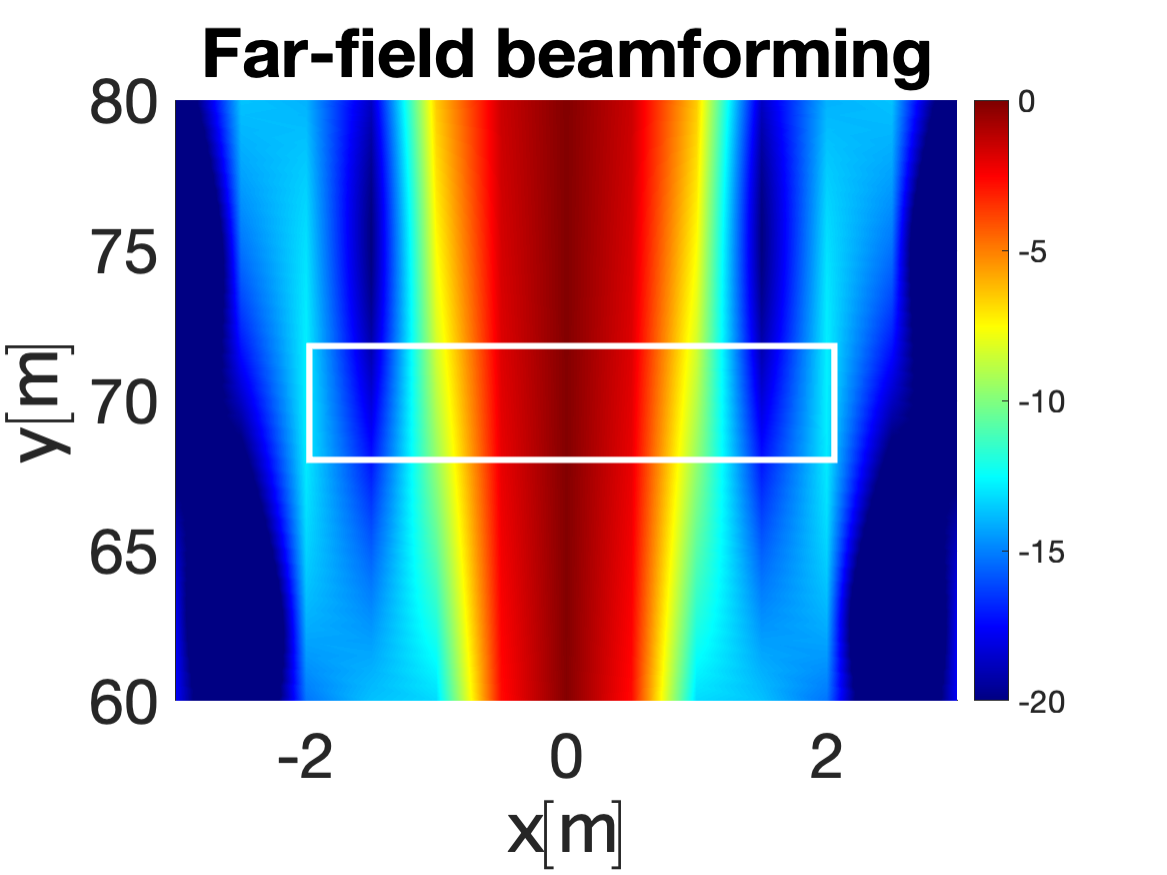}
    \caption{(F,S,$100$).}
    \label{fig: RIS designs_a}
\end{subfigure}
\begin{subfigure}{0.32\textwidth}
    \includegraphics[width=\textwidth,height=0.8\textwidth]{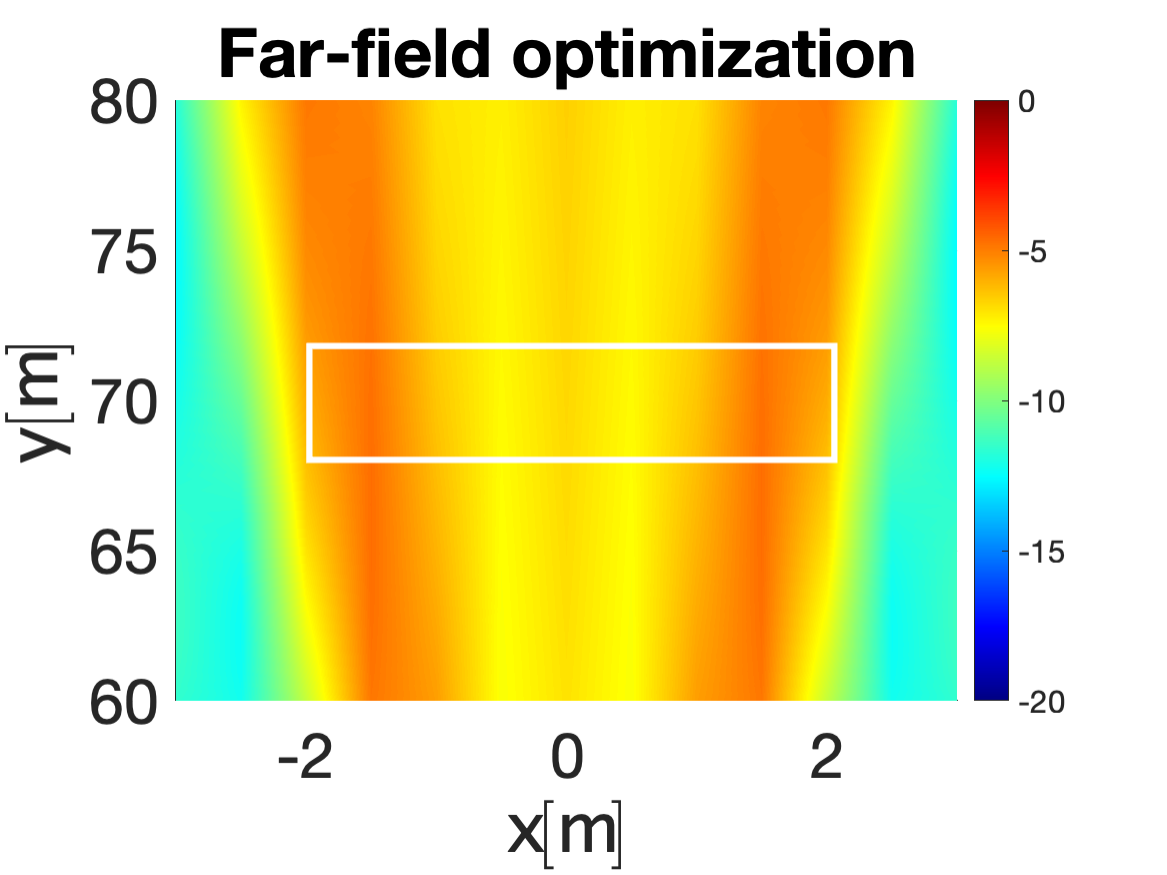}
    \caption{(F,R,$100$).}
    \label{fig: RIS designs_b}
\end{subfigure}
\begin{subfigure}{0.32\textwidth}
    \includegraphics[width=\textwidth,height=0.8\textwidth]{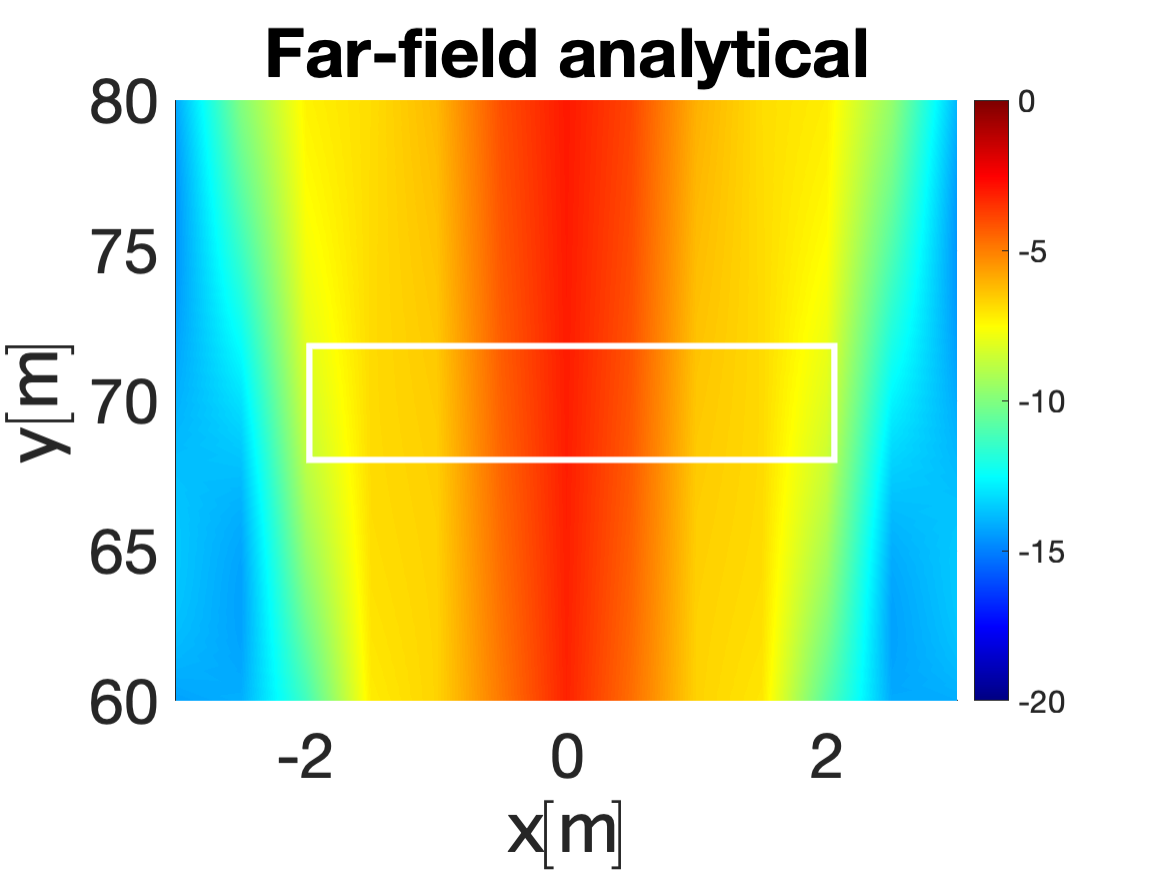}
    \caption{(F,R,$100$).}
    \label{fig: RIS designs_c}
\end{subfigure}
\begin{subfigure}{0.32\textwidth}
    \includegraphics[width=\textwidth,height=0.8\textwidth]{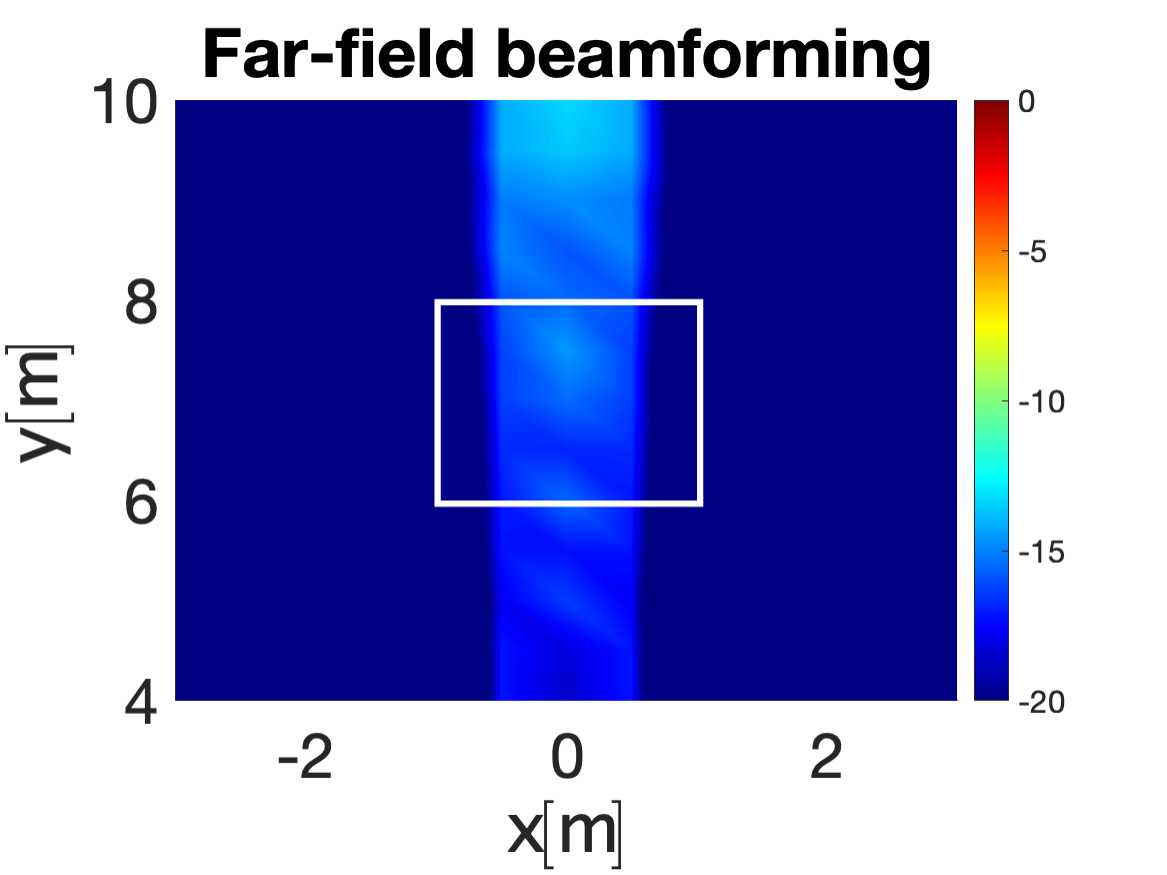}
    \caption{(N,S,$300$).}
    \label{fig: RIS designs_d}
\end{subfigure}
\begin{subfigure}{0.32\textwidth}
   \includegraphics[width=\textwidth,height=0.8\textwidth]{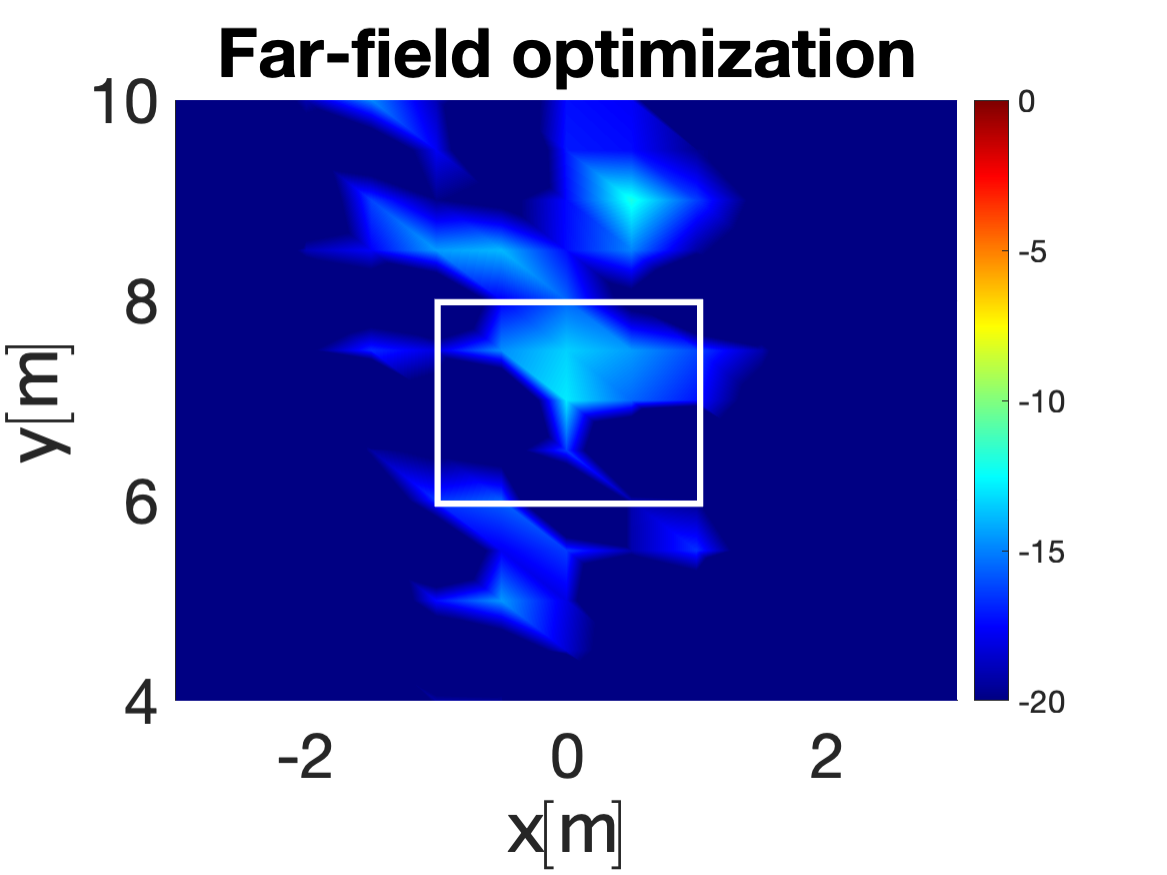}
    \caption{(N,R,$300$).}
    \label{fig: RIS designs_e}
\end{subfigure}
\begin{subfigure}{0.32\textwidth}
    \includegraphics[width=\textwidth,height=0.8\textwidth]{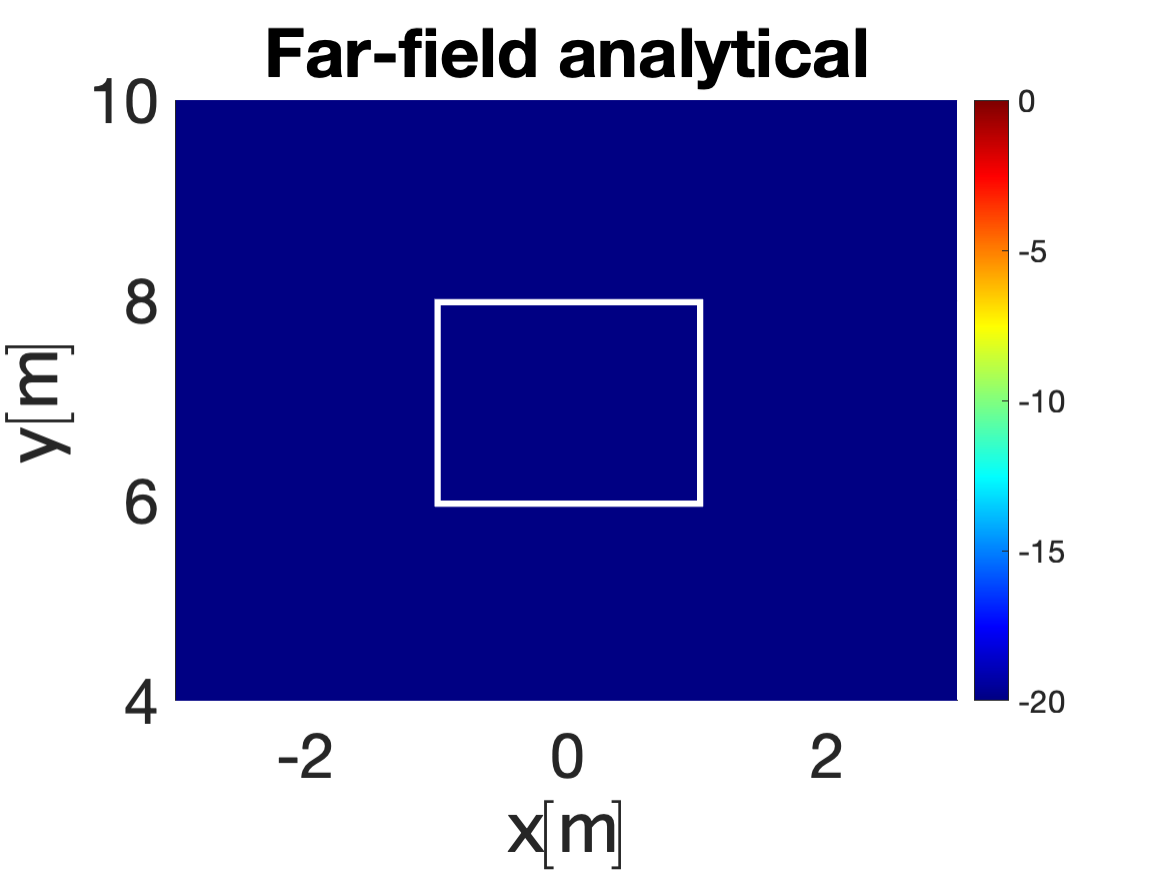}
    \caption{(N,R,$600$).}
    \label{fig: RIS designs_f}
\end{subfigure}
\begin{subfigure}{0.32\textwidth}
    \includegraphics[width=\textwidth,height=0.8\textwidth]{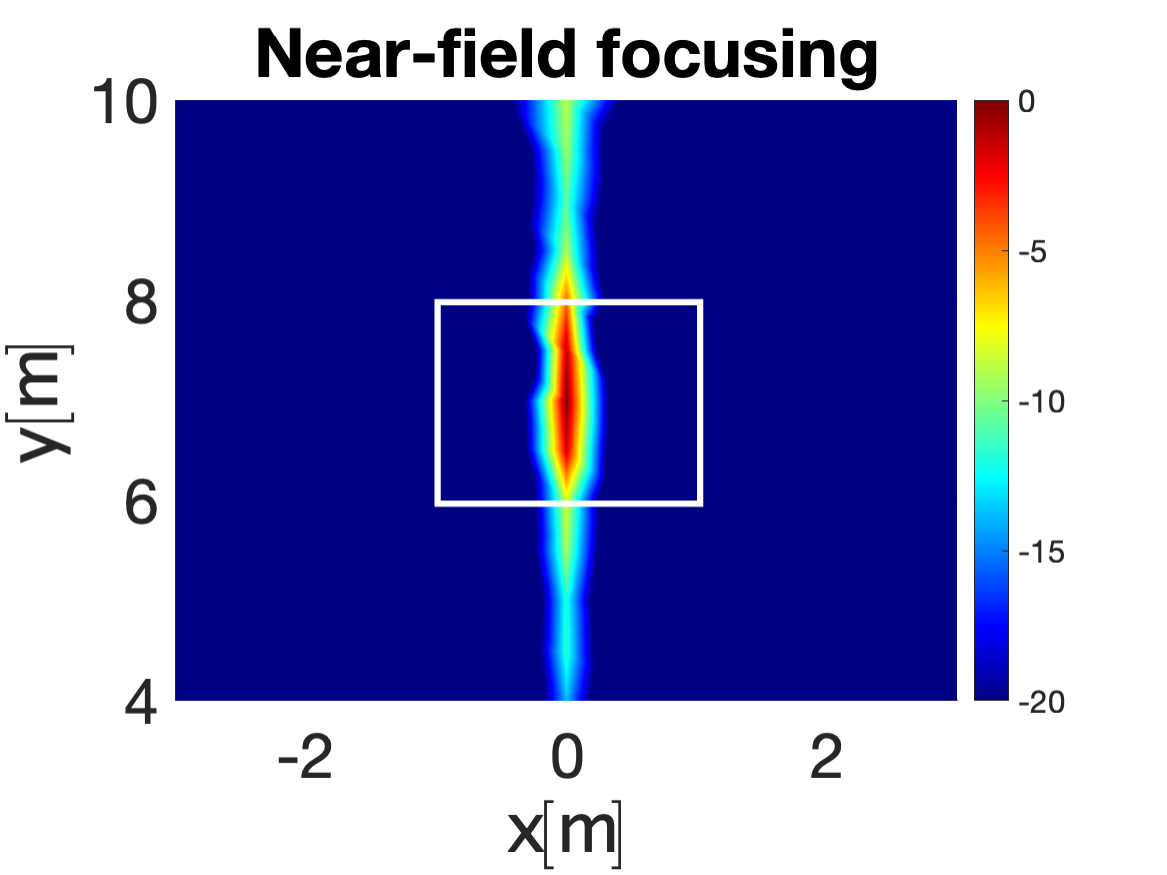}
    \caption{(N,S,$300$).}
    \label{fig: RIS designs_g}
\end{subfigure}
\begin{subfigure}{0.32\textwidth}
   \includegraphics[width=\textwidth,height=0.8\textwidth]{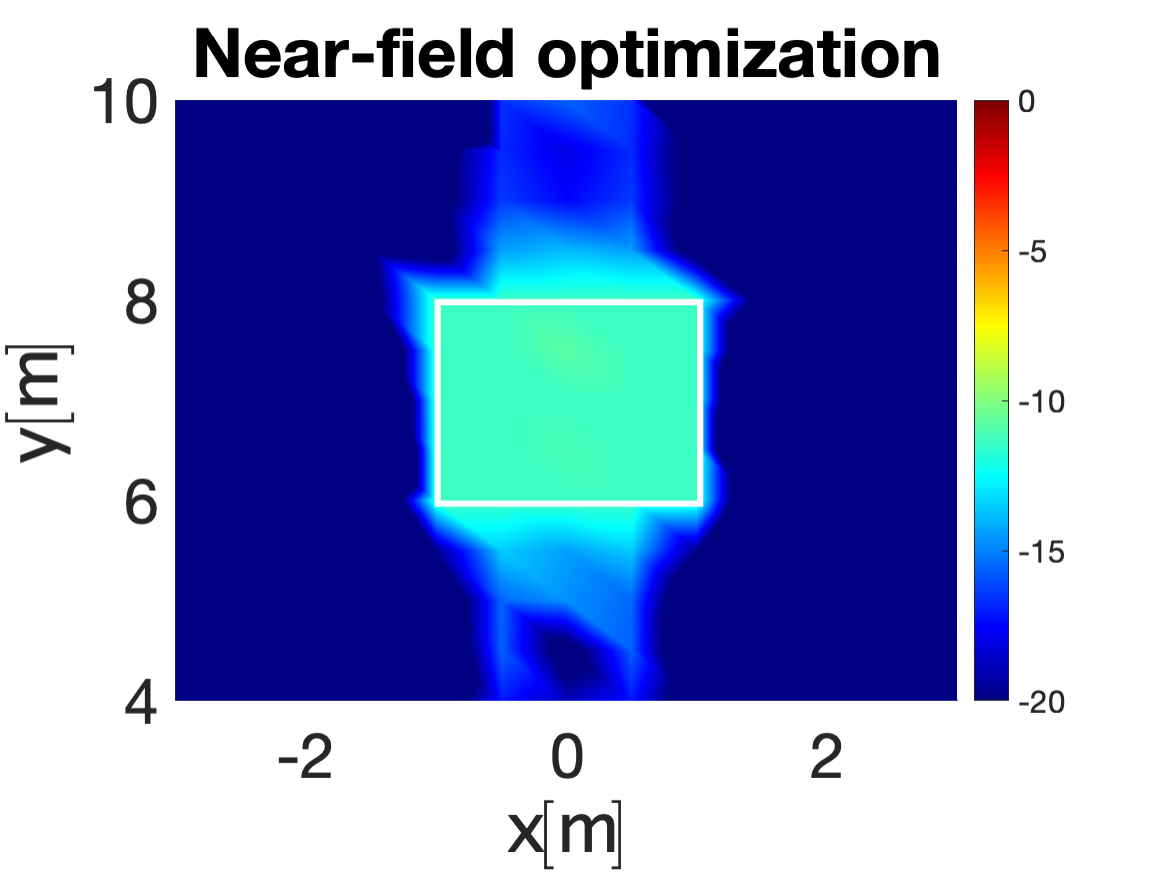}
    \caption{(N,R,$300$).}
    \label{fig: RIS designs_h}
\end{subfigure}
\begin{subfigure}{0.32\textwidth}
    \includegraphics[width=\textwidth,height=0.8\textwidth]{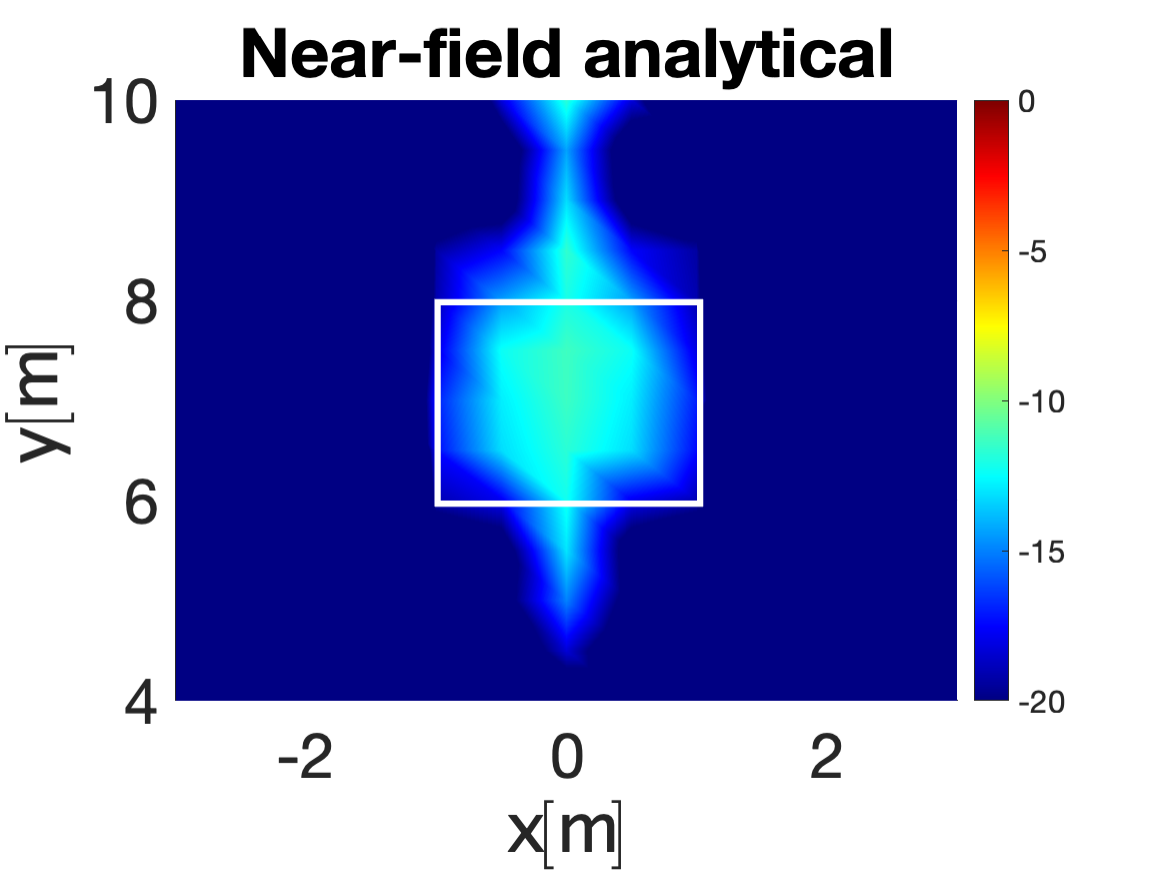}
    \caption{(N,R,$600$).}
    \label{fig: RIS designs_i}
\end{subfigure}
        
\caption{Normalized GRCS $\min\limits_{q\in\Qset} \bar{g}^2_\ris(q)$ (dB) for several far- and near-field scenarios. The first row includes results for the far-field regime when far-field designs are used; the middle row presents results for the near-field regime when far-field designs are adopted; and the bottom row demonstrates results for the near-field regime when near-field designs are employed. The abbreviation (A,B,C) used in the sub-captions is defined in the text.}
\label{fig: RIS designs}
\end{figure}


\textbf{Tunable Illumination:} We now investigate the performance of the proposed designs for the RIS reflective beams for tunable illumination, where the tunability of the illumination size is needed for a robust design in the presence of estimation errors for the channel parameters or for reducing the overhead of RIS phase reconfiguration (see Section~\ref{Sec:design}). In Fig.~\ref{fig: RIS designs}, we plot the normalized GRCS in dB for several far- and near-field scenarios. For conciseness, we use the abbreviation (A,B,C), where A=\{F,N\} indicating far- (F) and near-field (N), respectively, B=\{S,R\} implying a single point (S) and a region (R) of desired illuminated angle/point pairs, respectively, and C=$N$ is the number of RIS unit cells. The boundary of the desired illuminated region is specified by white lines. In the top row, we show the normalized GRCS in the far field for $N=100$, $\bu_\bs=[30,80,0]$~m, and $\bu_\ilm=[0,70,0]$~m (i.e., $d_\FF=54$~m, $d_t=85$~m, and $d_r=70$~m). As can be seen from Figs.~\ref{fig: RIS designs}~(a)-(c), both the optimization-based and analytical solutions offer tunable beamwidths as compared to standard beamforming. However, the optimization-based solution leads to a more uniform illumination compared to the analytical solution. In the remaining subfigures, we consider a near-field scenario, where $N=300,600$, $\bu_\bs=[30,80,0]$~m, and $\bu_\ilm=[0,7,0]$~m (i.e., $d_\FF=482,1928$~m, $d_t=85$~m, and $d_r=7$~m). In the middle row, we plot the results when the \textit{design} is based on the far-field assumption. To this end, Figs.~\ref{fig: RIS designs}~(d)-(f) clearly reveal that, neglecting the wavefront curvature in the near-field, significantly deteriorates performance and good coverage of the desired illumination area is not achieved. The bottom row shows results based on the proposed near-field design. From Fig.~\ref{fig: RIS designs}~(g), we see that near-field beam focusing leads to a narrow illumination spot on the desired coverage area. In addition, Figs.~\ref{fig: RIS designs}~(h) and (i) show results based on the optimization-based and analytical RIS designs, respectively, when illumination wider than that provided by full focusing is required. Again, the optimization-based design leads to a more uniform illumination; however, this is at the cost of higher computational complexity. In contrast, the analytical solution is more scalable and can be used for very large RISs (e.g., the considered RIS for the analytical solution has twice the size).

\begin{figure}[t]
     \begin{subfigure}{1\textwidth}
     \centering
         \includegraphics[width=\textwidth]{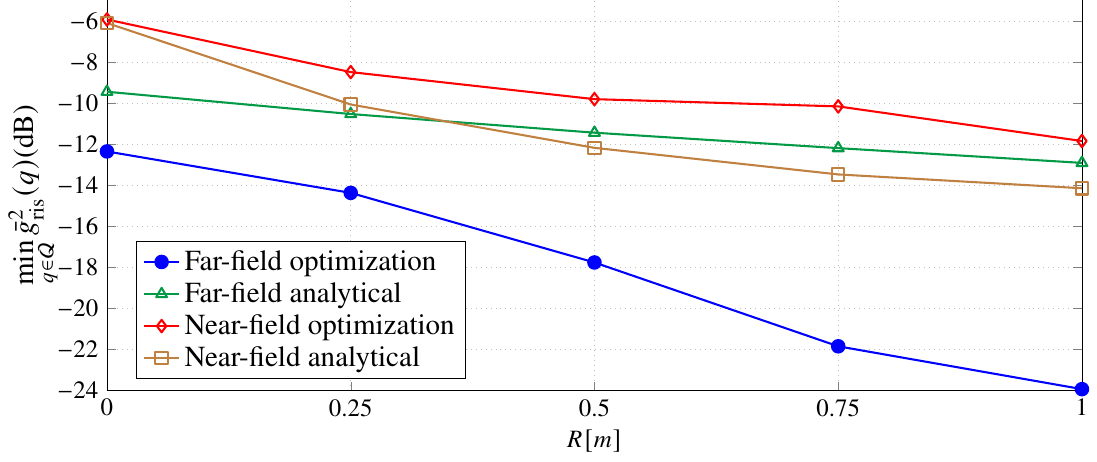}
         \caption{The coverage area is in the near-field regime.}
         \label{fig: error in receiver or transmitter_a}
   \end{subfigure}
        \begin{subfigure}{1\textwidth}
     \centering
         \includegraphics[width=\textwidth]{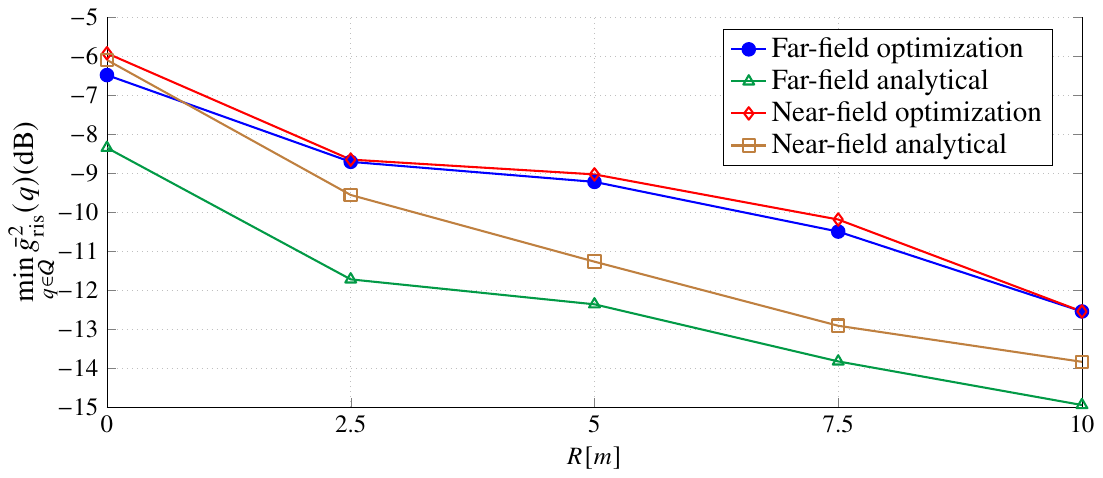}
         \caption{The coverage area is in the far-field regime.}
         \label{fig: error in receiver or transmitter_b}
   \end{subfigure}
   \caption{Normalized GRCS vs. the area length $R$ (i.e., a square area of size $R\times R$) for $N=100$, imperfect estimate of the BS location $\bu_\bs\in\{(\x,\y,\z):28~\text{m}\leq\x\leq 32~\text{m}, \y=80~\text{m}, \z=0\}$: (a) the near-field regime where $\bu_\ilm=[0,7,0]$~m, and (b) the far-field regime where $\bu_\ilm=[0,70,0]$~m.}
   \label{fig: error in receiver or transmitter}
   \end{figure}

In Fig.~\ref{fig: error in receiver or transmitter}, we quantitatively evaluate the performance of the analytical and optimization-based solutions in more detail. We assume that the estimation of the BS location is not perfect: $\bu_\bs\in\{(\x,\y,\z):28~\text{m}\leq\x\leq 32~\text{m}, \y=80~\text{m}, \z=0\}$, which implies an error of $\Delta\Theta=2.5^\circ$ for the estimation of the BS direction from the perspective of the RIS. The desired illuminated area is an $R\times R$ square region with center $\bu_\ilm=[0,7,0]$~m and $\bu_\ilm=[0,70,0]$~m in the near- and far-field regimes, respectively. The number of RIS unit cells is assumed to be $N=100$. In Fig.~\ref{fig: error in receiver or transmitter}~(a), we show the normalized GRCS as a function of the area length $R$ in the near field. As expected, the normalized GRCS decreases as the size of the desired illumination area increases, due to the distribution of the power over a larger region. In terms of performance, near-field optimization yields the highest normalized GRCS and far-field optimization yields the lowest normalized GRCS, due to the inaccurate adopted model. Interestingly, as $R$ increases, the far-field analytical solution attains a better performance than the near-field analytical solution. This is because, for large $R$, the effective size of the RIS responsible for a fixed size of the illumination area decreases, meaning that the far-field approximation for the RIS sub-area and the illumination sub-area become accurate. In this case, the difference between the far- and near-field analytical solutions is not about the accuracy of the underlying model, but the type of adopted mapping function, i.e., uniform angular discretization in \eqref{Eq:AlphaGamma} for the far field and uniform spatial discretization in \eqref{Eq:Mapping} for the near field. Since the choices of mapping functions in \eqref{Eq:AlphaGamma} and \eqref{Eq:Mapping} are heuristic, which of them performs better is scenario-dependent and not a priori known. In Fig.~\ref{fig: error in receiver or transmitter}~(b), we show the normalized GRCS as a function of area length $R$ in the far field. Here, the underlying models for both far- and near-field are valid. This figure reveals that the optimization-based methods outperform the analytical solutions. Moreover, the slight difference in performance for the far- and near-field optimization methods originates from the different discretization of the spaces of the BS location and illumination area (i.e., uniform angular discretization for the far field and uniform spatial discretization for the near field). Finally, for the considered setup, Fig.~\ref{fig: error in receiver or transmitter}~(b) suggests that the mapping function in \eqref{Eq:Mapping} is more effective than that in \eqref{Eq:AlphaGamma}, particularly for small $R$. 

\textbf{Impact of Scattering/Multipath:} A 3D simulation setup, as illustrated in Fig.~\ref{fig:system model}, in a multipath propagation environment is now considered. In particular, recalling that the RIS center is the origin of the coordinate system, we assume that the BS and MU are at heights $ \z= 5~\text{m}, -5~\text{m}$, respectively, i.e., $\bu_\bs=[30,80,5]$~m and $\bu_\ilm=[30,-5,-5]$~m. The RIS is a square surface and contains $N=100\times100=10000$ elements. implying that the near-field model is valid (i.e., $d_\FF=107~\text{m}, d_t=85.5~\text{m}, d_r=30.8$~m). We assume that a direct BS-MU link exists, which is however severely blocked by $-40$~dB. For the LOS components of the BS-RIS and RIS-MU links, we adopt the path loss model $h_{0}(d_0/d)^{\varepsilon}$ \cite{Jamali2023impact}, where $d$ is the link distance, $d_0$ is a reference distance at which the path loss is $h_{0}$, and $\varepsilon$ is the path loss exponent. Here, we adopt {$d_0=1$~m, $h_{0}=-61$~dB, and $\epsilon=2$}. For the BS-RIS and RIS-MU channels, non-LOS paths of both reflector and point-scatter type are considered. In particular, reflection from the ground is modeled as a perfect reflection with an average loss of $8$~dB w.r.t. the LOS path and an additional random Gaussian fluctuation~\cite{jaeckel2014quadriga,jaeckel2017quadriga}, where the ground is assumed to be at height $\z=-6$~m ($1$~m below the MU). For point-scattering, the location of the scattering object is generated randomly, where each scatter path itself contains $20$ sub-paths for emulating small-scale fading at each observation point. The overall power of the scattering paths is set according to the desired Rician factor, which is defined as the relative power of the LOS path, $c_0$, and the non-LOS paths, $c_v$, i.e., $K=\frac{|c_0|^2}{\Ex\{\sum_{v=1}^V |c_{v}|^2\}}$, where $V$ is number of non-LOS paths assumed six in this simulation (one ground reflection and $V-1$ point scatterers). Finally, we assume that the MU is subject to self-blockage in half of the space. With this assumption, the LOS link between MU and RIS is blocked with a probability of $50\%$ because of the human body; see \cite{dembele2021impact} for accurate models for the self-blockage probability.

We consider the following benchmark schemes. \textit{Benchmark~1:} This scheme employs full CSI and adapts the RIS unit-cell phase shifts to maximize the SNR \cite{Wu2021}. Therefore, it requires frequent CSI estimation according to the channel coherence time. \textit{Benchmark~2:} This scheme configures the RIS only based on the BS-RIS and RIS-MU LOS links, and hence, requires estimates of the BS and MU locations, which vary much more slowly than the channel CSI. Moreover, we show results for random phase shifts (leading to random reflection) and specular reflection as \textit{Benchmark 3} and \textit{Benchmark 4}, respectively. Finally, for the proposed scheme, we assume that the knowledge of both LOS links and $V_{\rm est}\leq V$ estimated non-LOS channel components is available, and the best path pair is selected for maximizing the SNR. The value of $V_{\rm est}$ can be related to the overhead required for identifying channel scatterers, i.e., the larger $V_{\rm est}$, the larger the corresponding estimation overhead. We adopt the near-field analytical solution for Benchmark~2 and the proposed scheme, respectively. 

\begin{figure}[t]
    \centering
\includegraphics[width=\textwidth]{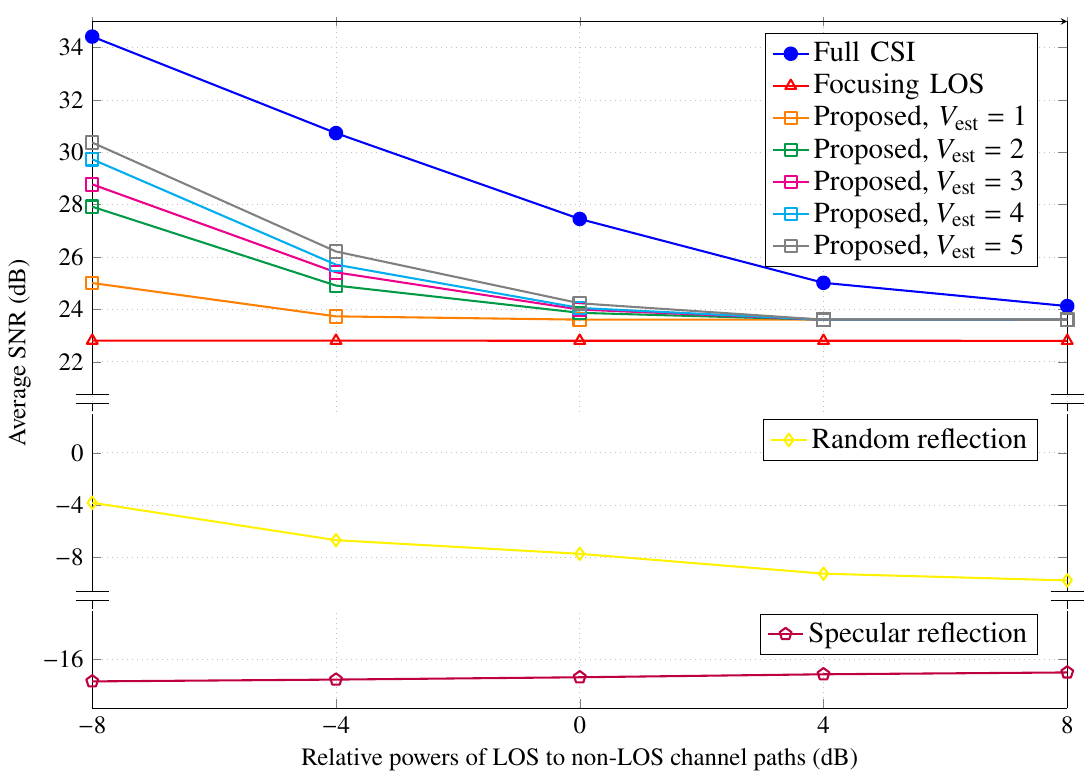}
    \caption{Average received SNR at the MU in dB vs. the relative power of the LOS path w.r.t. the non-LOS paths (dB).}
    \label{fig: 3D_model_simulation}
\end{figure}

In Fig. \ref{fig: 3D_model_simulation}, we plot the achievable SNR in dB as a function of the Ricean factor $K$ in dB. As can be seen from this figure, the full CSI scheme achieves the best performance at the cost of high CSI acquisition overhead. On the other hand, specular and random reflections, which do not exploit any CSI, lead to significantly lower achievable SNRs, which are several orders of magnitude smaller. Full focusing yields an almost constant SNR w.r.t. the Rician $K$-factor since, due to the narrow RIS beam, a negligible power is leaked to non-LOS paths making the achievable SNR of focusing independent of the power of the non-LOS components.  Interestingly, a considerable gain is achieved by the proposed scheme w.r.t. full focusing when accounting for non-LOS links, which often corresponds to reflection from the ground when $K$ is large enough. In fact, even for large Rician $K$-factors, an additional SNR gain is achieved by the proposed scheme w.r.t. full focusing due to the diversity of using non-LOS links in case of self-blockage.  Moreover, by accounting for multiple non-LOS links, an additional performance gain can be achieved when the Rician $K$-factor is a small number, but this gain ultimately saturates due to the negligible contribution of the weak point-source scatters.

\section{Conclusion}\label{sec:conclusions} 

This chapter presented a mathematical foundation for the modeling and design of passive and solely reflective RISs in the far- and near-field regimes. In particular, these regimes as well as the quadratic near-field sub-region were mathematically characterized, and end-to-end channel models for the far- and near-field regimes were introduced. Subsequently, two general approaches for RIS reflective beam design were presented, namely, one based on an optimization formulation and another analytical. The presented simulation results revealed that the optimization-based designs yield a higher quality RIS beam design compared to the analytical solutions. However, their computational complexity scales cubically with the number of RIS unit elements, and hence, becomes prohibitively expensive for large RISs, which motivates the use of analytical solutions in this case. Furthermore, using the proposed RIS designs, we showcased that, in sparse mmWave systems, transmitting along the dominant channel path leads to achieving higher SNR w.r.t. always focusing along the LOS path, mainly because of potential LOS blockage due to self-blockage, and the existence of strong nLOS components such as reflection from the ground. This performance gain is quite significant at low Rician $K$-factor values, although in this regime, there is still a considerable performance gap to the benchmark that assumes full CSI, thus, being associated with a high CSI acquisition overhead. Interestingly, as the value of the $K$-factor increases, the performances of the proposed scheme, LOS-focusing scheme, and full-CSI scheme converge. Therefore, compared to these benchmarks, the proposed scheme offers a favorable tradeoff between the achievable performance and the CSI acquisition overhead.


 \section*{Acknowledgements}
 Delbari and Jamali’s work was supported in part by the Deutsche Forschungsgemeinschaft (DFG, German Research Foundation) within the Collaborative Research Center MAKI (SFB 1053, Project-ID 210487104) and in part by the LOEWE initiative (Hesse, Germany) within the emergenCITY center. Alexandropoulos' work was supported by the Smart Networks and Services Joint Undertaking (SNS JU) project TERRAMETA under the European Union's Horizon Europe research and innovation programme under Grant Agreement no. 101097101, including top-up funding by UK Research and Innovation (UKRI) under the UK government's Horizon Europe funding guarantee. Schober’s work was funded by the German Research Foundation (DFG) under project SCHO 831/15-1 and the BMBF under the program of ``Souverän. Digital. Vernetzt.'' joint project 6G-RIC (Project-ID 16KISK023).

\bibliographystyle{IEEEtran}
\bibliography{References}

\end{document}